\documentclass{aa}  
\usepackage{graphicx}
\usepackage{txfonts}
\usepackage{natbib}
\usepackage{hyperref}
\usepackage{multirow}
\begin{document}

   \title{What excites the optical emission in X-ray-selected galaxies? \thanks{The catalogue of X-ray galaxies selected from SDSS is only available in electronic form at the CDS via anonymous ftp to cdsarc.u-strasbg.fr (130.79.128.5) or via \href{http://cdsweb.u-strasbg.fr/cgi-bin/qcat?J/A+A/}{http://cdsweb.u-strasbg.fr/cgi-bin/qcat?J/A+A/}}}

   \author{N.G. Pulatova
          \inst{1,2},
          H.-W. Rix
        \inst{1},
        A.V. Tugay
        \inst{3},
         L.V. Zadorozhna
         \inst{3,4,5},
                R. Seeburger
        \inst{1}
        \and M. Demianenko
        \inst{1}
                   }

   \institute{Max-Planck-Institute for Astronomy, K\"onigstuhl 17, D-69117 Heidelberg, Germany\\
              \email{pulatova@mpia.de}
              \and
         Main Astronomical Observatory of the National Academy of Sciences of Ukraine, Akademika Zabolotnoho str. 27, Kyiv 03143, Ukraine       \and
                Department of physics, Taras Shevchenko Kyiv National University,  Glushkova 2 ave., building 1, Kyiv 03680, Ukraine
  \and
        Niels Bohr Institute, Experimental Particle Physics, Blegdamsvej 17, 2100 K\o benhavn  Copenhagen, Denmark
       \and     
             Jagiellonian University Faculty of Physics, Astronomy and Applied Computer Science
        ul. prof. Stanislawa Lojasiewicza 11, 30-348 Krakow \\
     %\email{c.ptolemy@hipparch.uheaven.space}
             %\thanks{The university of heaven temporarily does not                     accept e-mails}
             }

  % \date{Received September 15, 1996; accepted March 16, 1997}

% \abstract{}{}{}{}{} 

  \abstract{We present a study of $1347$ galaxies at $z<0.35$ with detected nuclear X-ray emission and optical emission line diagnostics in the Baldwin-Phillips-Terlevich (BPT) diagram. This sample was obtained by cross-matching the X-ray Multi-Mirror Mission Observatory - Newton (XMM-Newton) DR10 catalogue with Sloan Digital Sky Survey (SDSS) DR17 galaxies with well-measured line ratios.
  %\LEt{***Please check that I have retained your intended meaning.***}. 
  The distribution of these sources in the BPT diagram covers all three excitation regimes: Ionized Hydrogen (HII) regions  (23\%), `composites' (30\%), and Seyfert galaxies with 
  %\LEt{***Please spell out all acronyms the first time they appear in the paper, followed by the abbreviation in parentheses, both in the abstract and again in the main text. After that, please only use the abbreviation. See A and A language guide Section 5.2.4 www.aanda.org/language-editing***}
  the low ionization nuclear emission line regions (LINERs) (47\%). In contrast, the fraction of objects classified as active galactic
nuclei (AGN) in the SDSS subsample selected for cross-match with XMM-Newton is only 13\%. This fact illustrates that X-ray emission from galaxies commonly points towards the presence of AGN. Our data show, for the first time, a clear dependence of the BPT position on the ratio of the X-ray to $H\alpha$ fluxes. Sources dominated by X-ray emission lie in the Seyfert and LINER regimes of the BPT diagram. Most sources with a low X-ray-to-$H\alpha$-luminosity ratio,  $log_{10}(L_X/L_{H\alpha}) < 1.0$, lie in the HII regime. In our sample, there are even 45 galaxies  that have $L^{Star}_{XR}/L^{Total}_{Xray}>0.5$. In contrast, the positions of  the sample
members in the BPT diagram exhibit {no} dependence on the X-ray hardness ratio. Our finding suggests that the X-ray-to-$H\alpha$ ratio can help us to differentiate galaxies whose X-ray flux is dominated by an AGN  {from galaxies with} central X-ray binaries and other stellar X-ray sources.%\LEt{***Please check that I have retained your intended meaning.***} 

  }


   \maketitle
%
%-------------------------------------------------------------------

\section{Introduction}

   The origin of the `nuclear emission' in galaxies, which is seen across a wide range of the electromagnetic spectrum, has been the focus of astrophysical studies for decades.   \citet{Baldwinet1981}
   proposed a powerful way to understand what physical processes dominate gas ionisation based on the flux ratios
of forbidden optical lines in the interstellar medium.   This criterion relies on four reddening-insensitive line ratios:  $[O III ]/H\beta$, $[N II ]/H\alpha$, $[S II ]/H\alpha$, and $[O I ]/H\alpha$.

The 
%%%\LEt{***Please spell out all acronyms the first time they appear in the paper, followed by the abbreviation in parentheses, both in the abstract and again in the main text. After that, please only use the abbreviation. See A and A language guide Section 5.2.4 www.aanda.org/language-editing***}
Baldwin-Phillips-Terlevich (BPT) diagram was intended to separate two sources of ionisation: active galactic
nuclei (AGN) or accretion discs around the supermassive black hole (SMBH), and young, massive hot stars, that is the host galaxy component. \citet{Kauffmann2003} defined the extreme starburst line to make a semi-empirical fit to the outer bound of the 
%\LEt{***Please spell out all acronyms the first time they appear in the paper, followed by the abbreviation in parentheses, both in the abstract and again in the main text. After that, please only use the abbreviation. See A and A language guide Section 5.2.4 www.aanda.org/language-editing***}
Sloan Digital Sky Survey (SDSS) galaxy spectra. This outer bound defines the region where composite starburst–AGN objects are expected to lie on the diagnostic diagrams. Later, \citet{Kewleyet2006} proposed an empirical relation to separate Seyfert galaxies from low ionisation nuclear emission-line regions (LINERs). \citet{Kauffmann2003} defined these latter as regions with [OIII] and $H\alpha$ narrow-line region (NLR) emission-line luminosities in the range of $\sim 10^5–10^6 L\odot$. LINERs are found in galaxies of earlier Hubble type than Seyfert galaxies, and their nuclear continua are usually dominated by old stars. According to the BPT diagram, galaxies with emission lines can be separated into the following categories depending on the type of source of ionisation that dominates there: Seyfert, LINERs, and HII regions.

   \begin{figure}
   \centering
   \includegraphics[width=\columnwidth]{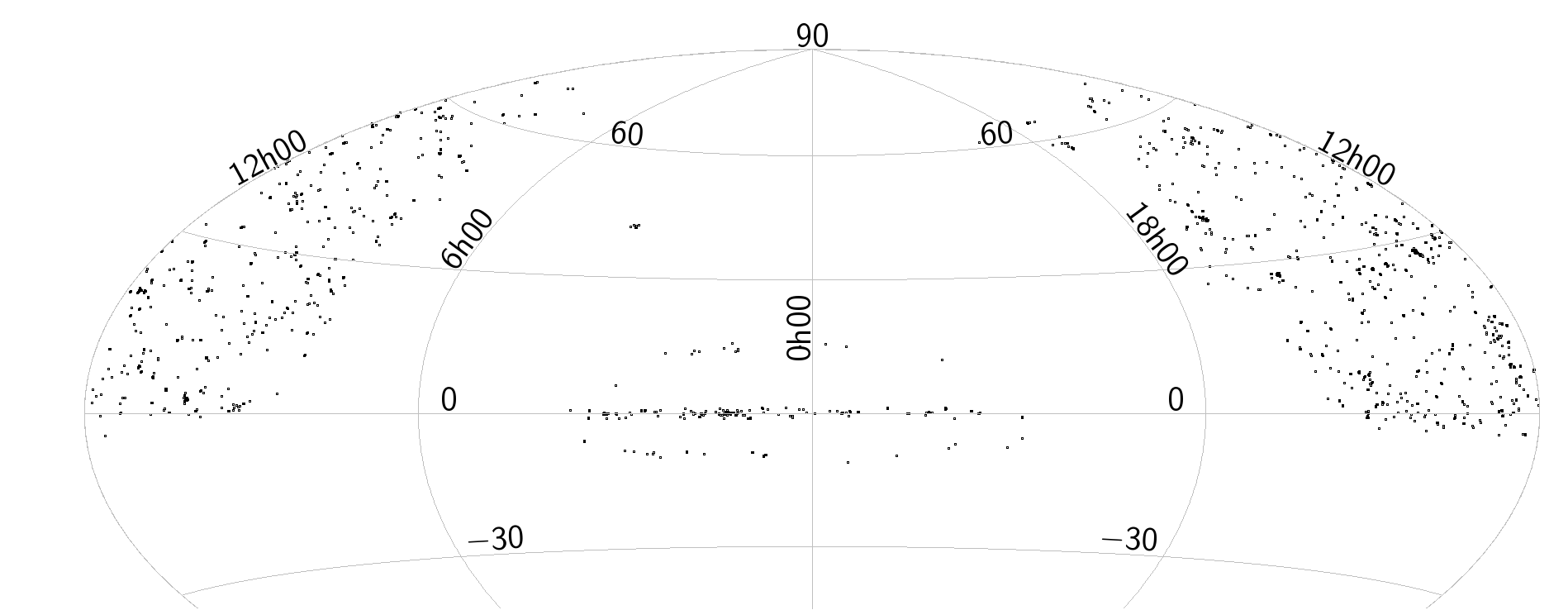}
  \caption{Distribution of the sample of 1347 X-ray-selected galaxies in equatorial coordinates.}
        \label{fig:RA-DE-disz}
    \end{figure}
    
\begin{figure}
\centering
        \includegraphics[width=\columnwidth]{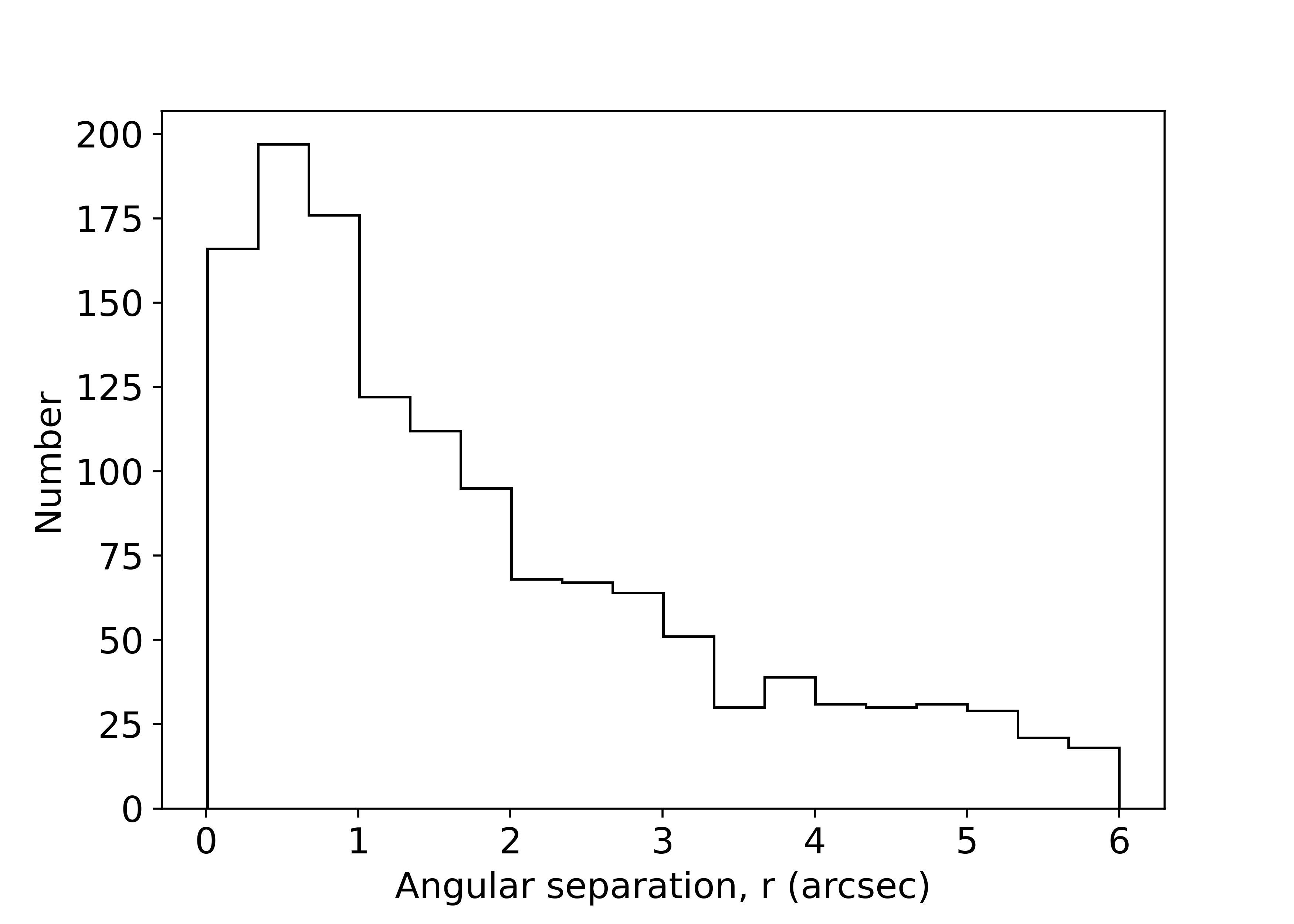}
        \caption{Distribution of 1347 SDSS X-ray galaxies in terms of the angular separation between the SDSS optical centre of the galaxy and the XMM-Newton X-ray source.}
        \label{fig:SDSS-X-ray-z-dist}
\end{figure}

\begin{figure}
\centering
        \includegraphics[width=\columnwidth]{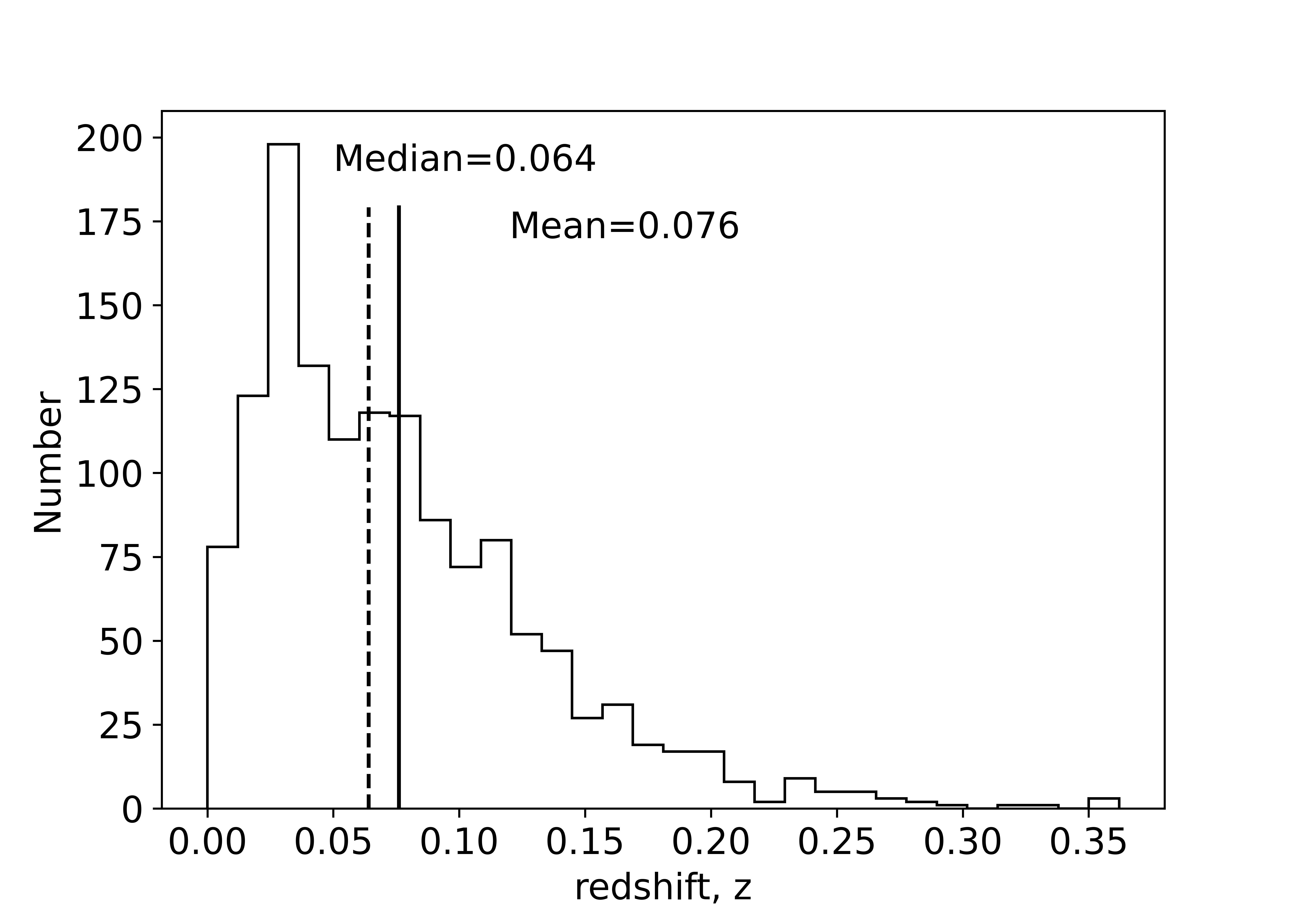}
        \caption{Distribution of 1347 SDSS galaxies with X-ray nuclear emission in terms of redshift, z .}
        \label{fig:SDSS-X-z}
\end{figure}

\begin{figure}
\centering
        \includegraphics[width=\columnwidth]{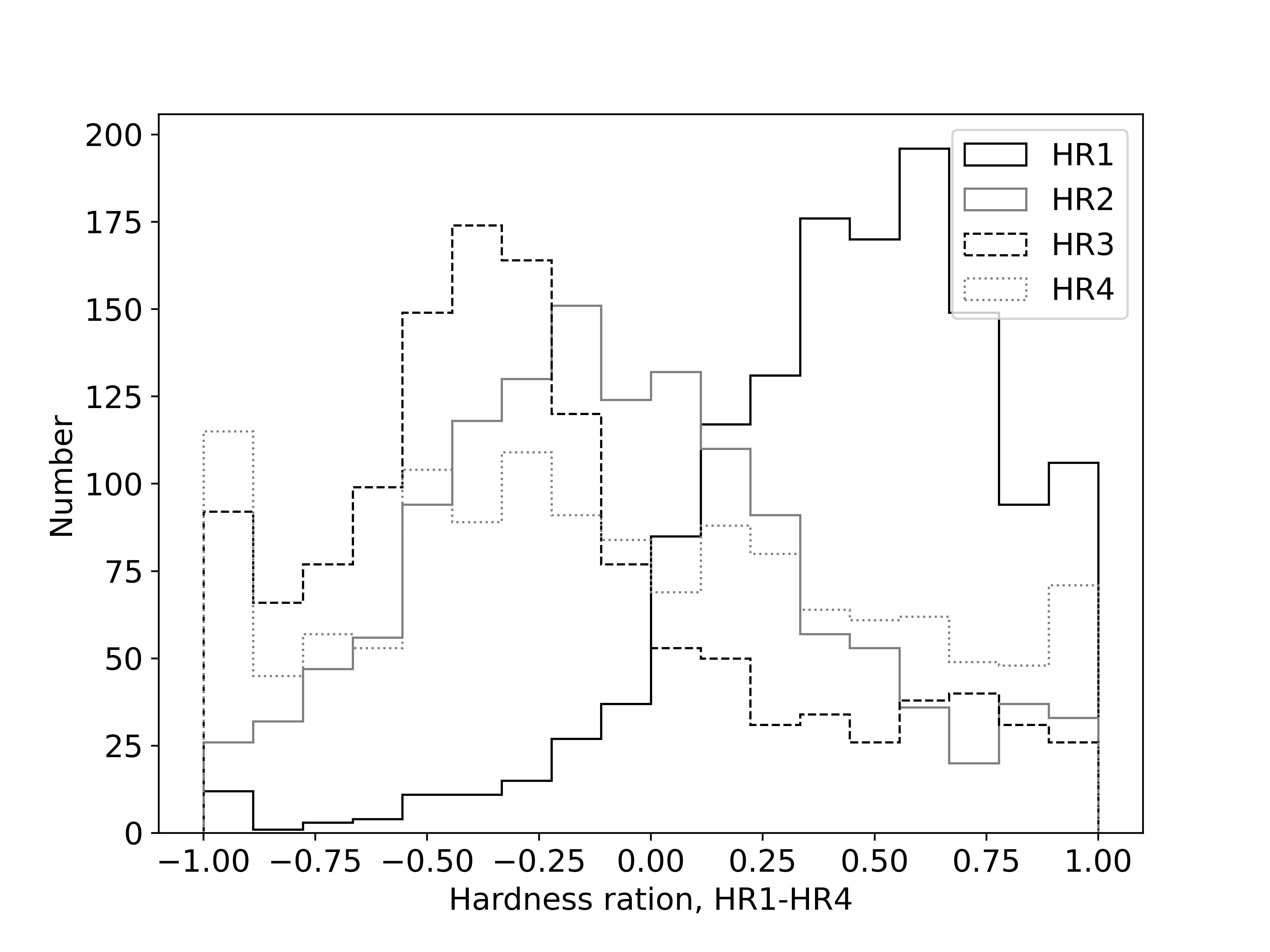}
        \caption{Distribution of 1347 SDSS X-ray-selected galaxies in terms of hardness ratio, HR1-HR4, according to data from the XMM-Newton catalogue.  $HRn=\frac{B_{n+1}-B_n}{B_{n+1}+B_n}$, where  $B$ denotes the count rates in energy band $n$, and $n$ takes a value from 1 to 4.}
        \label{fig:SDSS-X-HR}
\end{figure}

We investigated the X-ray emission from star forming regions and AGN separately. X-ray-selected galaxies can play a crucial role in our understanding of these sources: if the dominant X-ray emission inevitably comes from the accretion onto the SMBH, and if it is this process that
dominates the excitation of the optical emission lines, we should expect the optical line diagnostics of X-ray-selected galaxies to place them predominantly in the AGN regime in the Kewley version of the BPT diagram.
%Attempts have previously been made to tackle \textcolor[rgb]{0.984314,0.00784314,0.027451}{the issue we opened in our paper about X-ray-selected galaxies, namely AGNs}.\LEt{***The intended meaning here is unclear; please consider rewording and/or expanding.***} 
\citet{Torbaniuk2021} used a sample of X-ray-selected galaxies to quantify the level of AGN activity by applying multi-wavelength AGN-selection criteria (optical BPT diagrams, X-ray/optical ratio, etc.). The authors compared specific black hole accretion rates for AGN-dominated, quiescent galaxies and galaxies with active star formation. They concluded that the level of AGN activity decreased with cosmic time, and they found a scenario where both star formation and AGN activity are fueled by a common gas reservoir.

%--------------------------------------------------------------------
\section{SDSS galaxies with nuclear X-ray emission}

In this work, we study the X-ray emission and optical emission lines from X-ray-selected galaxies from SDSS Sky Survey DR17 \citep{2017AJ....154...28B}. To this end, we created a new sample of $1347$ galaxies at z<0.35 by cross-matching SDSS DR17 with X-ray Multi-Mirror Mission Observatory - Newton (XMM-Newton) DR10 \citep{webb20}.

\subsection[DATA]{Data}

In our work, we used data from the XMM-Newton observatory to study the X-ray emission of galaxies.
We chose 4XMM-Newton X-ray Source Catalogue
Data Release 10 (hereafter 4XMM-DR10), as it is one of the largest publicly available repositories of X-ray data from a single observatory \citep{webb20}. 
The 4XMM-DR10 database contains $849~991$ detections of    $575~158$ unique sources over $2.85\%$ of the sky. 
We used data from the PN camera, which has an angular resolution of 6.6 arcsec (full width at half maximum (FWHM)) in the energy range of 0.2 to 15 keV \citep{Struder2001}. We cross-matched 4XMM-DR10 with 204 895 SDSS-selected galaxies using the `Sky' algorithm  from the TOPCAT software with a separation of $6.0 ''$. For more information, see \href{https://www.g-vo.org/pmwiki/uploads/VOWorkshop/topcat.pdf}{TOPCAT Tutorial}, and Sect.~2.2). The sample of 204 895 SDSS galaxies was selected using the following criteria:

1. Galaxies with spectra with fractional flux uncertainties of $\le 33$\% or 0.13~dex \citep{Bolton2012A}

2. Narrow-line galaxies with $\sigma_{line}<200~km/s$. This criterion corresponds to the SDSS limit for narrow-line galaxies. Type 1 AGN galaxies with broad lines ($\sigma_{line}>200~km/s$) cannot be plotted correctly on the BPT diagram \citep{Bolton2012A}

3. These criteria were applied to emission lines $[O III],~[O I]~\lambda \lambda 5007, 6300$, $[N II]~\lambda \lambda 6548, 6584,$ $[S II ]~\lambda \lambda 6717, 6731$ doublets,  $H\alpha$, $H\beta$

\begin{figure*}
\centering
        \includegraphics[width=2.1\columnwidth]{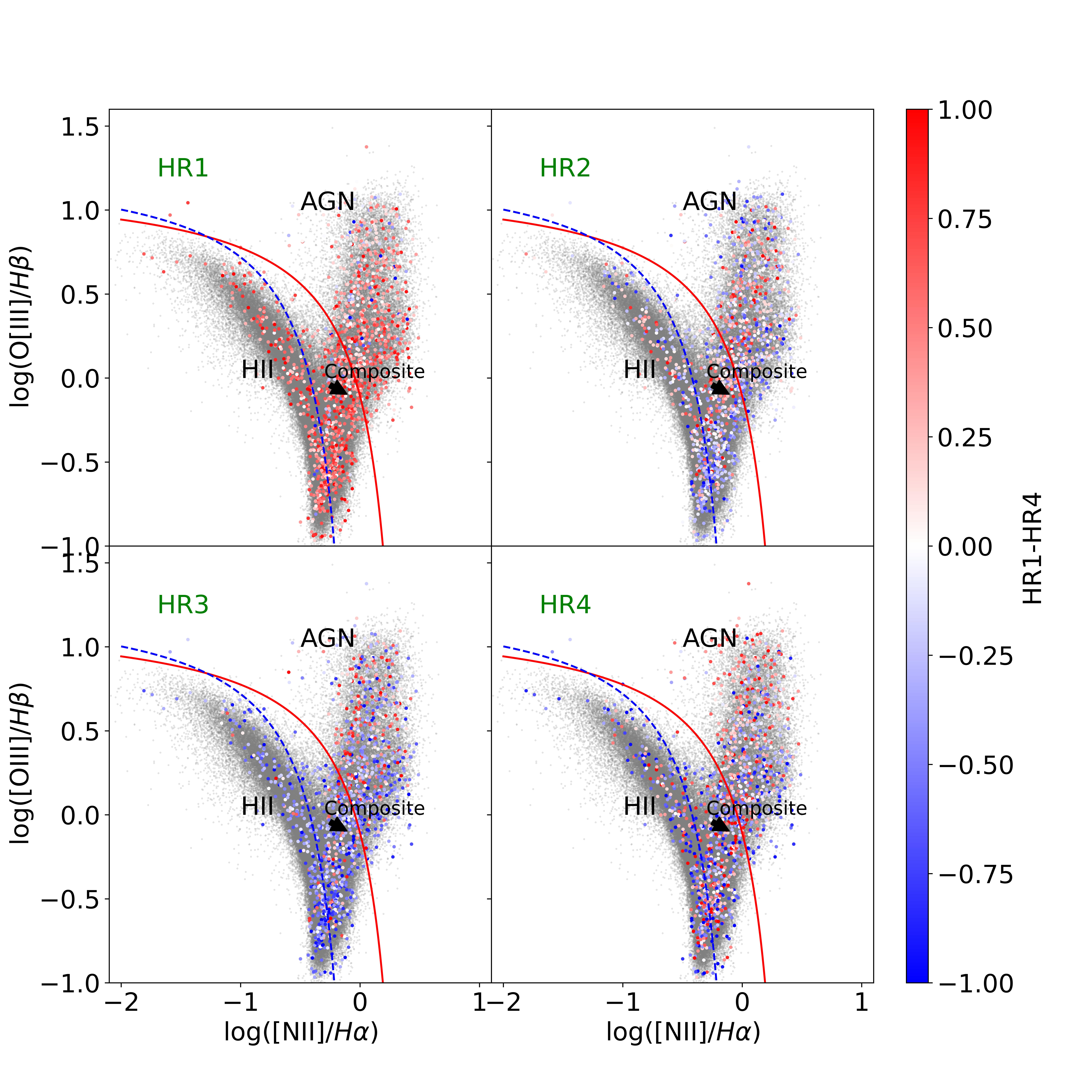}
        \caption{Position of 1347 SDSS X-ray-selected galaxies on the $[NII]/H{\alpha}$BPT diagram. The grey background circles show 204~895 SDSS-DR17 galaxies that satisfy the conditions for optical spectra $\sigma_{\mathrm{log}f} \le 0.13$ and $\sigma_{line}<200~km/s$. Empirical lines (dashed blue and solid red) that separate LINERs from Seyfert galaxies are from \citet{Kewleyet2006}. The colour code represents the hardness ratios among the HR1, HR2, HR3, and HR4 bands.}
        \label{fig:BPT-HR1-HR4}
\end{figure*}

\begin{figure*}
\centering
        \includegraphics[width=2.1\columnwidth]{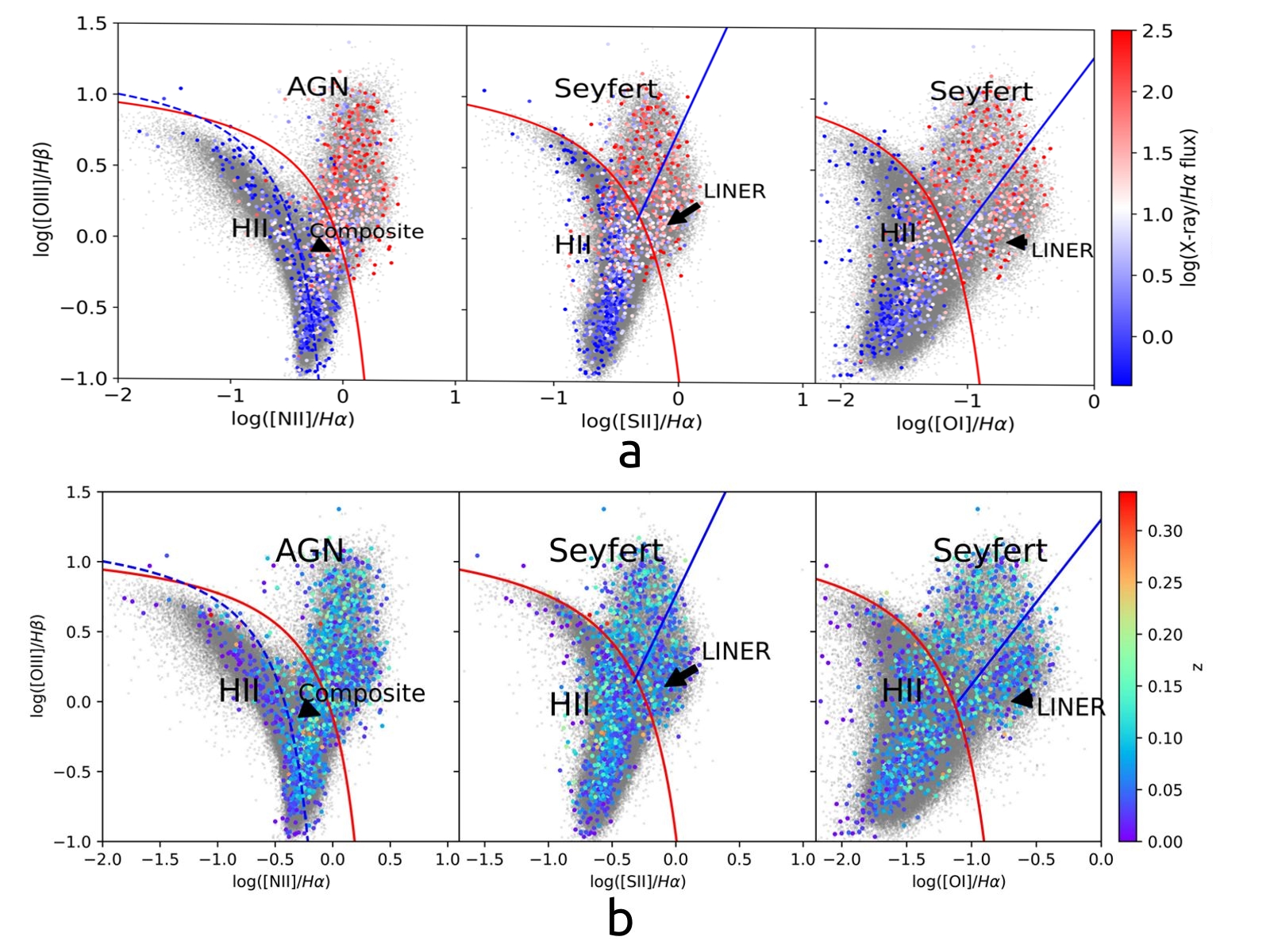}
        \caption{Distribution on the BPT diagram \citep{Baldwinet1981} of $1347$ SDSS X-ray-selected galaxies. Empirical lines (dashed blue and solid red) that separate LINERs from Seyfert galaxies are from \citet{Kewleyet2006}. The grey background circles show all the $204~895$ SDSS-DR17 galaxies that satisfy our basic optical quality cuts: all line fluxes have fractional uncertainties of $<0.13$~dex (33\%) and are `narrow' $\sigma_{line}<200~km/s$. The colour bar represents X-ray/H$\alpha$ flux ratio (6a) and redshift z (6b) and corresponds only to $1347$ SDSS X-ray-selected galaxies. }
        \label{fig:BPT-DR17}
\end{figure*}

 We also cross-matched the X-ray catalogue with the full SDSS sample and then applied the quality cuts after the matching. We ended up with nearly the same sample, which implies that source confusion (matching the wrong optical source to the X-ray source) is not a significant issue.

 \subsection[general properties]{Sample of X-ray galaxies with emission lines and its general properties}
Our cross-match result is a sample of $1364$ emission line galaxies with nuclear X-ray emission. Some of these matched sources could be incorrect because the closest match is not necessarily the counterpart. In our previous work, we carried out manual inspection of X-ray and optical sources from the SIMBAD and XMM-IRAP databases \citep{zador21}. In the present study, we quantified the potential contamination by shifting the positions of the X-ray sources by random amounts that are larger than the source-matching radius. After this, we again matched `shifted' X-ray sources with SDSS DR17. As a result, we only got 51 false matches, which corresponds to 3.7 \%. We used the optical emission-line fluxes from the SDSS DR17 data release to find the positions of X-ray-selected galaxies on the BPT diagrams. This allows us to determine the source of excitation in these galactic nuclei. After visual inspection of the BPT diagram (see Figure~\ref{fig:False-gal-BPT}), we found that some galaxies lie below the LINER position for most of the objects from the sample. We separated these galaxies from the primary sample and reviewed every individual galaxy (for more details, please see~the Appendix). After careful checking, we removed  17 galaxies from the primary sample. The last two galaxies, SDSS J015308.29+010150.6 and SDSS J130340.81+534323.6, are located at the end of the LINER region. We obtained a final sample that contains $1347$ X-ray-selected galaxies at $z<0.35$ that covers the northern sky.  This number corresponds to $0.6\%$ of the SDSS galaxies with good optical spectra and $0.2\%$ of the XMM-Newton X-ray sources, respectively.

For each galaxy, we included the following parameters: Right Ascension ($deg$), Declination ($deg$), redshift (z), optical emission line flux with errors: $[O III], [O I] \lambda 5007, 6300$, $[N II ] \lambda \lambda 6548, 6584, [S II ] \lambda \lambda 6717, 6731$ doublets,  $H\alpha$, $H\beta$ in   $10^{-17} erg \cdot s^{-1} \cdot cm^{-2} \cdot $\AA{}$^{-1}$, X-ray flux in the total photon energy band with errors ($0.2 - 12 keV$) from the XMM-DR10 catalogue ($erg/cm^2 \cdot s^{-1}$), hardness ratio (HR), and the angular separation between the position of the centre of a galaxy and the X-ray source. 

Figures~\ref{fig:RA-DE-disz}-\ref{fig:SDSS-X-z} present the general properties of the sample. Figure~\ref{fig:RA-DE-disz} shows the distribution of $1347$ galaxies with nuclear X-ray emission in equatorial coordinates. From this distribution, we see that SDSS X-ray-selected galaxies cover the northern sky, and follow the distribution of SDSS galaxies. The equator region with higher spatial density corresponds to Stripe 82 \citep{LaMassa2013, LaMassa2016}. SDSS observed this region intensively during 2005, 2006, and 2007. The zone of avoidance in the northern sky corresponds to the Milky Way galaxy. The distribution of X-ray-selected galaxies in equatorial coordinates is homogeneous over the northern sky. The distance between the X-ray source and the optical centre of the galaxy is less than 1 arcsec apart, for more than half of X-ray-selected galaxies
%\LEt{***The intended meaning here is unclear; please consider rewording}
(see Figure~\ref{fig:SDSS-X-ray-z-dist}). The number of X-ray galaxies rapidly decreases with angular separation from 1 arcsec to 3.5 arcsecs from $n=125$ at $r=1$ arcsec to $n=35$ at $r=3.5$ arcsec, and after $r=3.5$ arcsec and then come to a plateau. We plot the distribution of $1347$ SDSS galaxies with a  nuclear X-ray source as a function of redshift in Fig.~\ref{fig:SDSS-X-z}. The minimum and the maximum values are $\approx 0$ and $0.362$, respectively. For galaxies located at higher redshift ($z>0.35$), $H_\alpha$ optical line will be outside the SDSS cutoff wavelength of 9200\AA. For $z > 1.14$, $H_\beta$ is redshifted out of the wavelength range. The distribution of SDSS X-ray-selected galaxies peaks at $z \sim 0.03$, and then the number of galaxies decreases with increasing z. The mean and median values for the distribution as a function of redshift are 0.076 and 0.064, respectively.

Figure~\ref{fig:SDSS-X-HR} presents the distribution of the SDSS X-ray-selected
galaxies as a function of HR, showing HR1-HR4. The HR is the equivalent of a photometric color index, and is calculated as a normalized difference of the exposure corrected counts in two energy bands.
%\LEt{***The intended meaning here is unclear; please consider rewording and/or expanding.***} 

For HR1, the commonly used bands are 0.2 - 0.5 keV and  0.5 - 1.0  keV. For HR2, the bands 0.5 - 1.0  keV and 1.0 - 2.0  keV are used. For  HR3,       1.0 - 2.0 keV and 2.0 - 4.5  keV are used. Finally, for HR4, 2.0 - 4.5  keV and 4.5 - 12.0 keV are used. 
For more information about the hardness ratio, readers can refer to the XMM-Newton tools website \footnote[2]{https://xmm-tools.cosmos.esa.int/external/sas/current/doc/emldetect/node3.html}. From the distribution in the BPT diagram of X-ray-selected galaxies with the 
%\LEt{***Please spell out all acronyms the first time they appear in the paper, followed by the abbreviation in parentheses, both in the abstract and again in the main text. After that, please only use the abbreviation. See A and A language guide Section 5.2.4 www.aanda.org/language-editing***}
HR colour code, we conclude that there is no apparent correlation between the optical type (star formation (SF), AGN, composite) of X-ray galaxies and their HR. However, at the same time, we find a correlation between X-ray-to-$H\alpha$ flux ratio and optical type. This could be due to the location of the $H\alpha$ emission line (accretion disc), while X-ray emission comes from the corona.

From the general parameters of the sample, we conclude that most galaxies from the sample belong to the local Universe, and the sample can be considered homogeneous. A benefit of using this SDSS X-ray-selected galaxy sample is that it covers almost all of the northern sky, and is therefore representative of X-ray galaxies in the nearby Universe.
%\LEt{***Please check that I have retained your intended meaning.***} 

\section{Results}
\label{sec:BPT}

In this section, we show how we investigated the position of X-ray-selected galaxies in the BPT diagram as a function of two quantities: the X-ray hardness ratio, HR1-4, and the X-ray-to-$H\alpha$ flux ratio. Figure~\ref{fig:BPT-HR1-HR4} displays the distribution of the 1347 X-ray-selected galaxies on the [NII]/$H\alpha$ version of the BPT diagram. The data points are colour coded according to X-ray hardness ratios (HR1-HR4). Figure ~\ref{fig:BPT-DR17}  presents the full range of BPT diagrams ([NII]/$H\alpha$,  [SII]/$H\alpha$, and  [OI]/$H\alpha$ as the x-axes). The colour code indicates the X-ray-to-H$\alpha$ flux ratio (Fig. 6a) and the redshift $z$ (Fig. 6b).
We also include a comparison with 204,895 galaxies from the SDSS-DR17 dataset in both figures. These galaxies meet the criteria of $\sigma_{\mathrm{log}f} \le 0.13$  for all lines and $\sigma_{line}<200 km/s$; they are plotted as a grey background in the diagrams.

\begin{itemize}
\item We confirm that X-ray-selected galaxies can lie in all regimes of the BPT diagram: Seyfert 25\%, LINERs 22\%, composite (30\%), and  HII (23\%). For the sample of $204~895$ SDSS galaxies with emission lines (plotted as a grey background),  AGNs make up only 13\%, which is consistent with \cite{Kewleyet2006}. 
This fact has two immediate implications: X-ray-selected emission-line galaxies have a much higher ($3.5\times$) incidence of AGN than the parent sample of galaxies with narrow emission lines; and the majority of X-ray-detected galaxies with narrow emission lines (at least in our sample) are not AGN dominated according to their optical emission lines.

\item From the distribution in the BPT diagram of X-ray-selected galaxies with the HR colour code, we conclude that there is no apparent correlation with HR.
\item We checked how many galaxies from our sample have the X-ray luminosity 
%\LEt{***The intended meaning here is unclear; please consider rewording and/or expanding.***}} 
$L_X>3 \cdot 10^{42}$ erg/s, and can be classified as AGN according to \citep{Torbaniuk2021}. According to this classification, we find 214 AGNs among 1347 X-ray-selected galaxies (16\%). The percentage of X-ray AGNs grows from the star forming galaxy category to the Seyfert optical category: 6  from 316 (1.8\%) star forming galaxies, 
52 from 405 (12.8\%) composite galaxies, and 176 from 626 (28\%)  Seyfert galaxies with LINERs. This means that X-ray AGNs ($L_X>3 \cdot 10^{42}$ erg/s) can be found in all optical categories. However, the highest percentage of X-ray AGNs is found in Seyfert galaxies and LINERs. This result shows that X-ray luminosity is a necessary but insufficient condition for identifying a source as an AGN.
\item We show, for the first time, that the X-ray-to-H$\alpha$ flux ratio is a good predictor of the BPT regime in which a galaxy located: almost all the objects with the highest X-ray-to-$H\alpha$ ratios, $log_{10}({L_X/L_H}_\alpha) >1.0,$ are located in the AGN regime. Almost all the galaxies with low ratios, $log_{10}({L_X/L_H}_\alpha) <1.0, $ are located in the H~II regime. 

\end{itemize}

The distinct pattern in the  colours of the data points (reflecting $log_{10} (L_X/L_H\alpha)$ in Fig.~\ref{fig:BPT-DR17}a (top) reflects the power of the $log_{10} (L_X/L_H\alpha)$ flux ratio as a
predictor of the excitation mechanism. 
The bottom panel of Fig.~\ref{fig:BPT-DR17} explores whether aperture corrections or the distance to the object play a significant role. The absence of a redshift dependence in the BPT position implies that such effects are insignificant.

\begin{figure}
    \centering
        \includegraphics[width=\columnwidth]{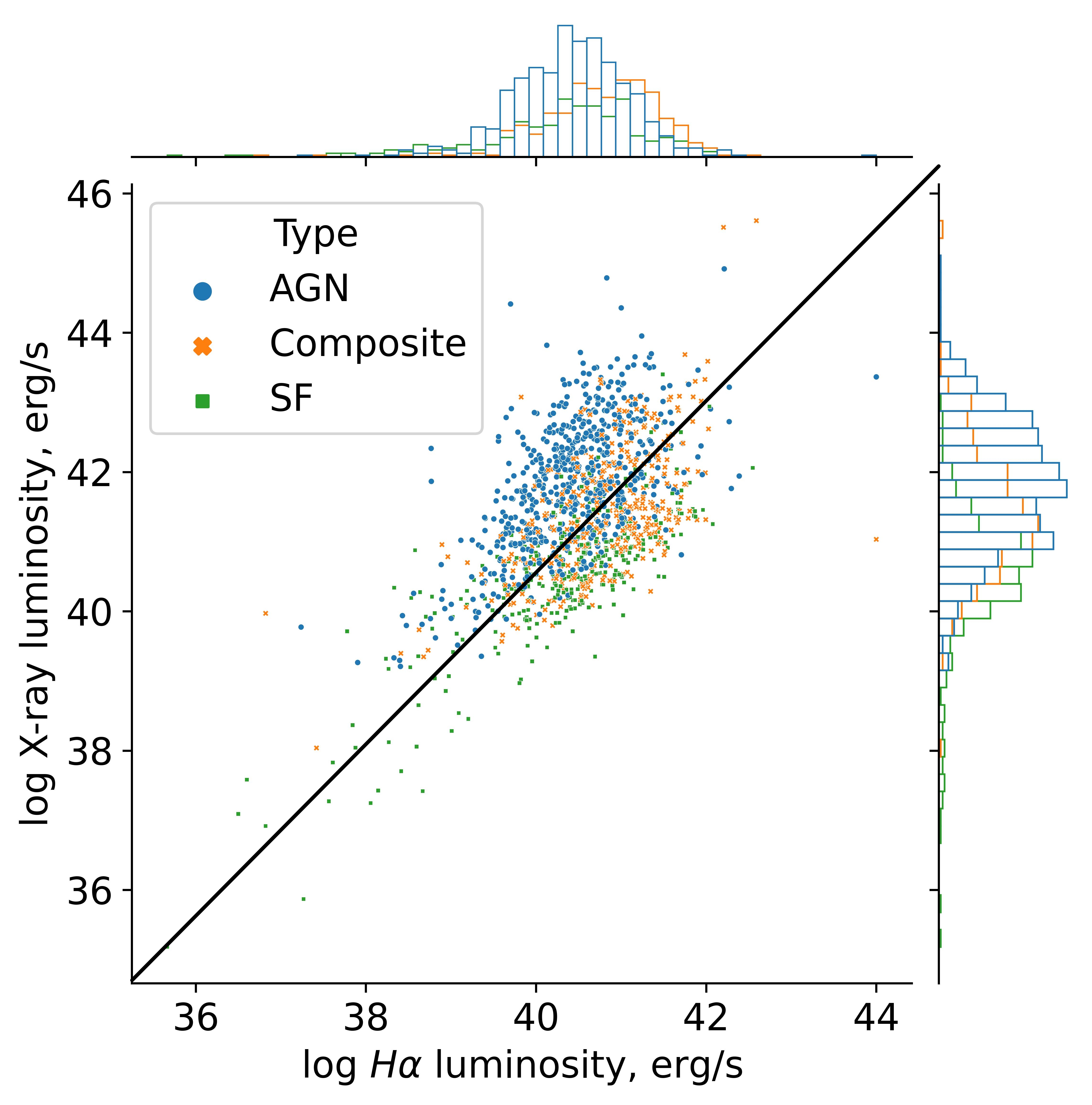}
        \caption{Positions of 1347 SDSS X-ray selected galaxies on X-ray versus $H\alpha$ luminosities plane. X-ray luminosity is in the range 0.2-12 keV and in $erg\cdot s^{-1}$, $H\alpha$ luminosity is in $erg\cdot s^{-1}\cdot$. Type of ionisation (AGNs, Composite, SF) corresponds to the position of X-ray galaxy on [OIII]/$H\beta$ versus [NII]/$H\alpha$ BPT diagram. AGNs are marked with blue circles, Composites - with orange crosses, and SF - with green squares.}    \label{fig:X-ray-H-al}
\end{figure}
To detect the type of ionisation that influences the X-ray and H$\alpha$ luminosities, we plot the distribution of SDSS X-ray-selected galaxies in the X-ray versus $H\alpha$ plane in Fig.~\ref{fig:X-ray-H-al}. The plot is similar to Fig.~\ref{fig:BPT-DR17} but adds additional information about the distribution of X-ray and $H\alpha$ luminosities for different types of excitation mechanisms in galaxies.  
In Fig.~\ref{fig:X-ray-H-al}, we separate the whole sample of $1347$ SDSS X-ray-selected galaxies into three groups (AGN, composite galaxies, and star-forming galaxies) according to their position on the $[NII]/H\alpha$ diagram (source of ionisation). These three groups are colour coded in Fig.~\ref{fig:X-ray-H-al} . 

\begin{figure*}
    \centering
        \includegraphics[width=2.1\columnwidth]{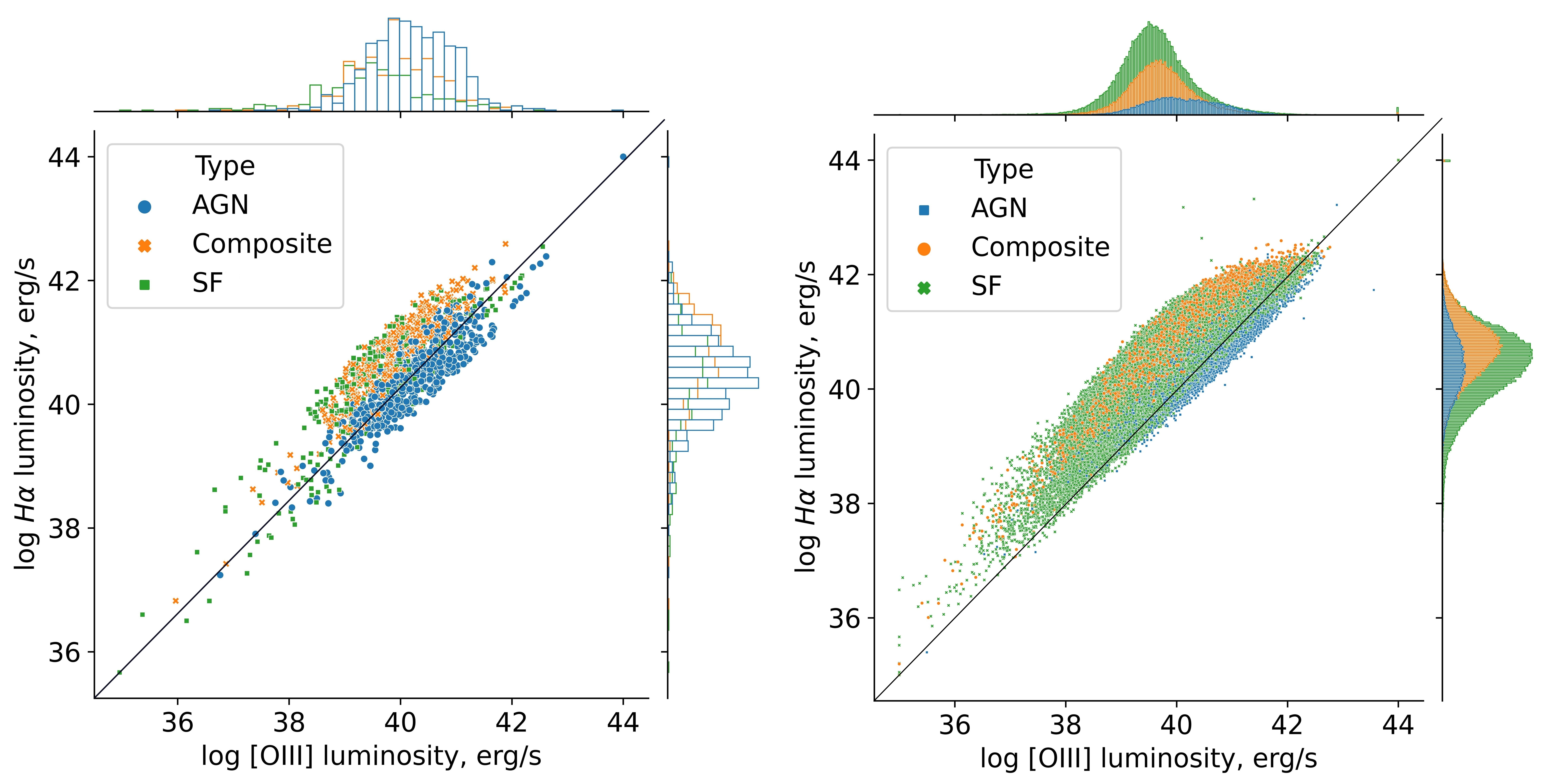}
\caption{Distribution on the $log_{10} [OIII]$ versus $log_{10} H\alpha$ plane of SDSS X-ray selected galaxies (left) and 204 895 SDSS galaxies (right). Type of ionisation (AGNs, Composite, SF) corresponds to the position of X-ray galaxy on [OIII]/$H\beta$ versus [NII]/$H\alpha$ BPT diagram.}
\label{fig:H-alpha_Lum_comb}
\end{figure*}

From Fig. ~\ref{fig:X-ray-H-al}, we see that the peaks of the distribution of the X-ray-selected galaxies correspond to different types of ionising sources. The first peak in $H\alpha$ luminosity at  $3\cdot 10^{40}$ corresponds to AGN, and the second two belong to composite and star-forming galaxies. The distribution of $H\alpha$ luminosity for composite and star-forming galaxies shows flat peaks of about $10^{41}~erg/s$. 
The peak in the $H\alpha$ luminosity distribution for AGNs is one order less than 
%\LEt{***The intended meaning here is unclear; please consider rewording and/or expanding.***}
for composite and star-forming galaxies.
The opposite situation is seen for X-ray luminosity. From Fig.~\ref{fig:X-ray-H-al}, we see that the maximum of the distribution of X-ray-selected AGNs in the energy range of 0.2-12 keV corresponds to  $10^{42}~erg/s$, while for star-forming and composite galaxies, the maximum of the distribution matches $10^{41}~~erg/s$. 
%It is seen from the X-ray versus $H\alpha$ distribution. 
The distribution of X-ray-selected AGNs is almost symmetrical to the black line in the centre of Figure~\ref{fig:X-ray-H-al}, and star-forming
and composite X-ray-selected galaxies are located mostly below the black line.  AGNs are mostly brighter in X-rays than star-forming
and composite galaxies. 

\begin{figure*}
        
        \includegraphics[width=2.1\columnwidth]{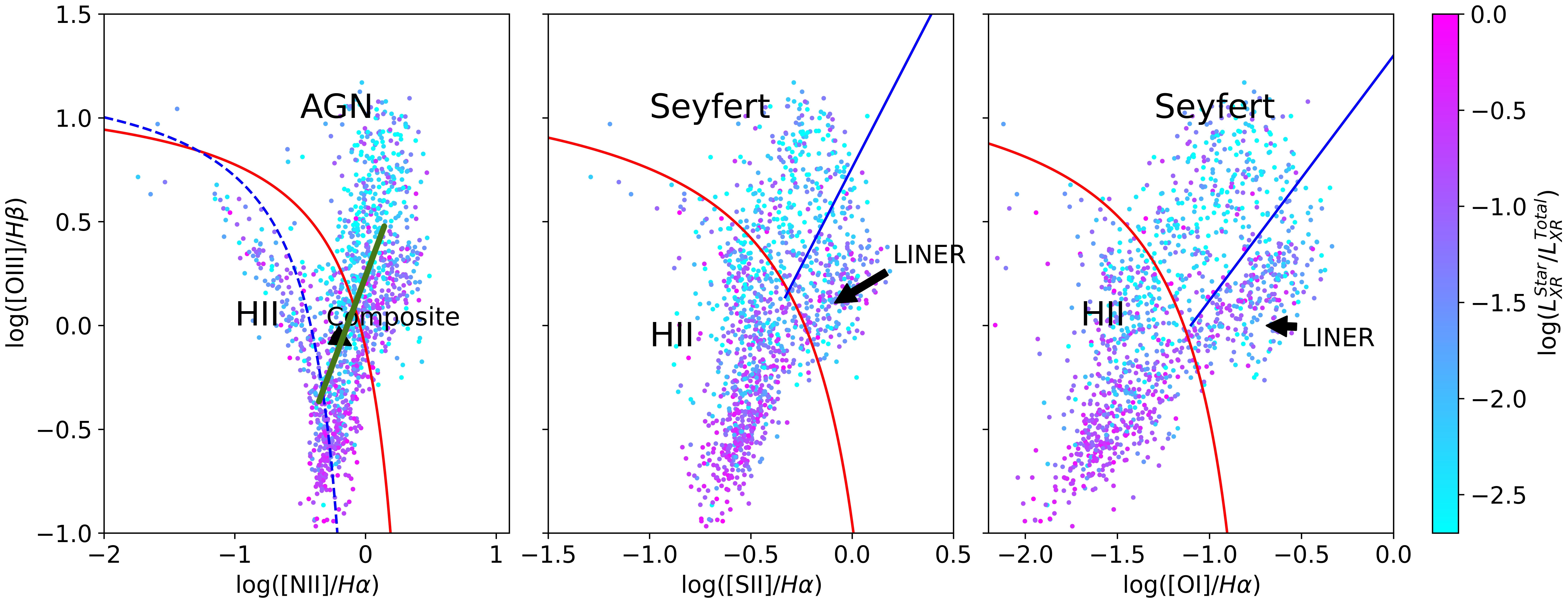}
    \caption{BPT diagram for 1262 SDSS X-ray selected galaxies. Empirical lines (dashed blue and solid red) that separate LINERs from Seyfert galaxies are from \citet{Kewleyet2006}.  The relation between X-ray luminosity from the stellar component and the total X-ray luminosity from the XMM-Newton observations is shown with colour coding.}
    \label{fig:figure1}
\end{figure*}

\begin{figure*}
        
        \includegraphics[width=2.1\columnwidth]{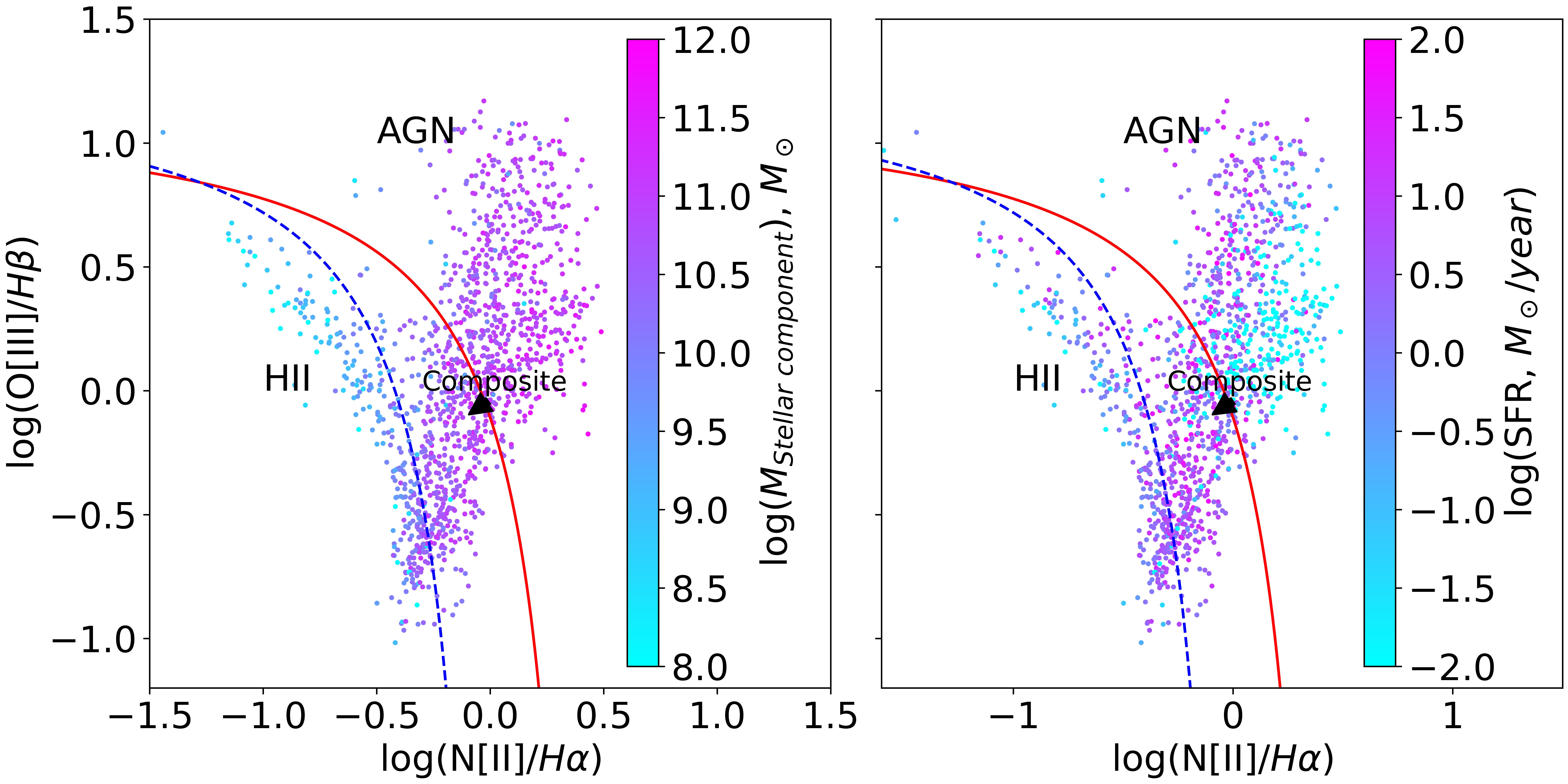}
    \caption{BPT diagram for 1262 SDSS X-ray-selected galaxies. The empirical lines (dashed blue and solid red) that separate LINERs from Seyfert galaxies are from \citet{Kewleyet2006}. The mass of the stellar component (left) and the SFR (right) from \citet{Chang2015} are shown with colour coding in the respective panels. 
    }
    \label{fig:figure}
\end{figure*}

The distribution on the $log_{10} [OIII]$ versus $log_{10} H\alpha$ luminosity plane of SDSS X-ray-selected galaxies (left) and 204 895 SDSS galaxies (right) (See Fig. ~\ref{fig:H-alpha_Lum_comb}) shows the same general behaviour. The brightest in [OIII] luminosity are AGNs in both samples with peak values of about $10^{40}~ erg/s$. Composite galaxies then follow, with peak values of about $8 \cdot 10^{39}~ erg/s$. The weakest of the three optical categories is the star-forming galaxies, with a peak in [OIII] luminosity corresponding to $5 \cdot 10^{39}~ erg/s$.
A different situation is seen for  $log_{10} H\alpha$ luminosity. The brightest are composite galaxies with peak values of about $10^{41}~ erg/s$ for both samples. The distributions of star-forming galaxies and AGNs both show peaks at $5 \cdot 10^{40}~ erg/s.$
The main difference between the two plots is that  AGN dominate the distribution of X-ray-selected galaxies, while star-forming galaxies dominate the SDSS sample.
%\LEt{***Please check that I have retained your intended meaning.***}

\begin{table*}[]
        \centering
        \caption{Distribution of $1347$ SDSS X-ray-selected galaxies and 204~895 SDSS galaxies (the last column), according to the their position on the BPT diagram}
        \label{tab:BPT-content}
        \begin{tabular}{lccc|r} % four columns, alignment for each
                \hline
                Type of ionisation & $[NII]/H\alpha$ & $[SII]/H\alpha$ & $[OI]/H\alpha$ &  $[NII]/H\alpha$ 204~895 SDSS galaxies\\
                \hline
                Star-forming      &       316 (23\%)      & \multirow{2}{*}{797   (59\%)}  &       \multirow{2}{*}{699     (52\%)} & 116~139 (57\%)\\
                Composite      &       405 (30\%)      &   &           & 62~039 (30\%)\\
                Seyfert &  333 (25\%) & 291 (22\%)      & 347  (26\%)   &  \multirow{2}{*}{26~527 (13\%)}\\
                LINERs  & 293 (22\%) &  259     (19\%) &        301  (22\%)     & \\
                
                \hline
        \end{tabular}
\end{table*}

%\subsection{X-ray emission from star-forming galaxies}
\label{sec:maths} % used for referring to this section from elsewhere
We calculated the part of X-ray emission from the stellar component in SDSS X-ray-selected galaxies. To this end, we used the method described in \citet{Lehmer2010}:

\begin{equation}
    L_{HX}^{gal}=\alpha \cdot M_*+\beta \cdot SFR,
        \label{eq:X}
\end{equation}
where $\alpha = (9.05\pm 0.37) \cdot 10^{28}~erg \cdot s^{-1} \cdot M^{-1}_\odot$ and $\beta = (1.62\pm 0.22) \cdot 10^{39}~erg \cdot yr \cdot s^{-1} \cdot M_\odot^{-1} $

$L_{HX}^{gal}$ traces the combined emission from high-mass X-ray binaries (HMXBs) and low-mass X-ray binaries \citep{Lehmer2010}. This empirical equation was obtained based on high-resolution Chandra observations.

For the $M_*$ and SFR in Eq.~(\ref{eq:X}), we used data from \citet{Chang2015}. 

For 1262 SDSS X-ray-selected galaxies, we obtained the relation between X-ray emission from X-ray binaries and the total X-ray emission from XMM-Newton data.
%\LEt{***Please check that I have retained your intended meaning.***} 
The results are plotted in Fig.~\ref{fig:figure1}.

The AGN component dominates the X-ray emission for almost the entire sample $(L^{Star}_{XR}<0.5\cdot L^{Total}_{XR})$, except for 45 objects that are preferentially located at the lower end of the BPT. These galaxies could be pure X-ray AGN-free galaxies \citep{Tugay2011}. The AGN dominance is stronger for Seyferts than for LINERS and HII regions. To investigate this question 
%\LEt{***The intended meaning here is unclear; please consider rewording and/or expanding.***}
further, we plotted the star-formation rate (SFR) and masses of the stellar component on the BPT diagram, using a color code (see Fig.~\ref{fig:figure}). By separating the stellar X-ray components, we see that LINERS have higher stellar masses and much lower SFRs than Seyferts. In the HII region,  objects with a low SFR also have low masses.

\section{Discussion and conclusion}

It is commonly accepted that most X-ray emission in the Universe originates from AGN
%\LEt{*** up to here in the article, the plural of AGN has been denoted simply AGN. If the authors wish to use AGNs then please edit throughout for consistency, using "AGNs" whenever referring to more than one AGN.***},
but some amount could come from the stellar population of a host galaxy. For a detailed study of the physical mechanism inside AGNs and star-forming galaxies, separating stellar and AGN components in optical spectra is essential. Contributions to the emission-line spectrum from both star formation and the AGN are almost inevitable in many SDSS galaxies, given the relatively large projected aperture size of the fibres ($5.5$ kpc diameter at $z = 0.1$; \cite{Kauffmann2003}).

The connection between star formation and the central 
%\LEt{***Please spell out all acronyms the first time they appear in the paper, followed by the abbreviation in parentheses, both in the abstract and again in the main text. After that, please only use the abbreviation. See A and A language guide Section 5.2.4 www.aanda.org/language-editing***}
SMBH has been investigated in multiple studies. Stellar feedback from a young nuclear star cluster has been shown to influence the interstellar medium at the centres of AGN 
%\LEt{***Please check that I have retained your intended meaning.***}
\citep{Schartmann2009, Costa2014, Hopkins2016, Shangguan2020}. However, AGN located inside the galaxy can also influence star formation in the host galaxy. These effects are called negative and positive AGN feedback. Through such processes, AGN can suppress or trigger star formation in the host galaxy \citep{Croton2006, Ishibashi2012, Cresci2018, Stemo2020}.

The interdependence of the emission in X-rays and optical lines was discovered at the end of the 20th century.
For example, in luminous sources, strong correlations exist between $H\alpha$, $H\beta$, $[OIII]\lambda5007$ luminosities, and X-ray luminosities  \citep{Elvis1984, Ward1988, Mulchaey1994}. Many authors have used the X-ray-to-optical(continuum) flux ratio to identify AGN in their host galaxies \citep{Brandt2005, Fitriana2022, Torbaniuk2021}.  \citet{Brandt2005} list three AGN criteria: 

- an intrinsic X-ray luminosity, 
$L_{X,int} >= 3 \cdot 10^{42}~erg \cdot s^{-1}$;

- an X-ray-to-optical flux ratio of $log_{10}~( f_X / f_{opt} ) > -1;$

- an X-ray-to-IR flux ratio of $log_{10} ( f_X / f_{K_s} ) > -1.2$.

The interdependence of X-ray and optical flux in the r-band was revealed based on a study of a sample of 6181 X-ray sources from the Stripe-82X survey \citep{Ananna2017}. The authors claim that objects with $log_{10}~X/O>1$ and R-K(Vega magnitude)> 5 are candidates for heavily obscured AGNs. Our investigation partially confirms this result. Namely, we find that galaxies with $log_{10} (L_X/L_H\alpha)>1.0$ are linked to LINER or Seyfert-type nuclear activity.

A series of works is devoted to BPT diagnostic diagrams for X-ray AGNs.
 \citet{Birchallet2022} present a sample of $917$ X-ray-selected AGN within the local galaxy population. Similarly to the present work, these authors created their sample from galaxies listed in SDSS DR8 and 3XMM DR7. The BPT diagnostic diagram can be used to confirm an AGN inside a host galaxy. The BPT diagnostic is more effective at reliably identifying sources as AGN in higher-mass galaxies. These latter authors used this sample to calculate the incidence of AGN as a function of stellar mass and redshift. Their main finding is that the fraction of galaxies hosting AGN above a fixed specific accretion rate limit of $10^{-3.5}$ is constant (at $\sim 1$  \%) over stellar masses of $ 8 < \log_{10} M_*/M_\odot <$  12 and increases (from $\sim 1$\% to 10\%) with redshift.

 \cite{Torbaniuk2021} used the sample of X-ray-selected galaxies to quantify the level of AGN activity by applying multi-wavelength AGN-selection criteria (optical BPT-diagrams, X-ray/optical ratio, etc.). 
  
The present work is different because \cite{Torbaniuk2021} used flux in only one line from two doublets, [NII] $\lambda6583$ and [SII] $\lambda6717,$ to find the position of a galaxy on the BPT diagrams. In our work, we consider both doublet lines [NII] $\lambda\lambda$ 6548, 6584, [SII] $\lambda\lambda$ 6717, 6731, as was done in the original work of \cite{Baldwinet1981}. We restrict our sample to SDSS spectra whose fractional flux uncertainties in all lines are $\le 0.13$~dex. 

For a more detailed investigation of the relationship between X-ray and $H\alpha$ flux for AGNs and star-forming galaxies, we separated the sample of 1347 X-ray-selected galaxies  into
four groups according to their position on the BPT diagram and the type of ionisation that dominates; these are star-forming galaxies, composite galaxies, Seyfert galaxies, and LINERs (See Table~\ref{tab:BPT-content}). As Table~\ref{tab:BPT-content} shows, 23\% of the 1347 X-ray-selected galaxies belong to pure galaxies with star-forming regions. Seyfert galaxies and LINERs make up less than 50\% of all X-ray galaxies and can be divided into almost equal parts of about 20\% each. In the original work of \cite{Kewleyet2006}, the authors used the sample of $85~224$ galaxies, finding $63~893$ (75\%) star-forming galaxies, $2411$ (3\%) Seyferts, $6005$ (7\%) LINERs, and $5870$ (7\%) composite galaxies, as well as $7045$ (8\%)  galaxies of an ambiguous nature. We find that the fraction of galaxies with an AGN type of nuclear activity is significantly higher among X-ray-selected galaxies  (47\%) compared to that of the total sample of $204~895$ SDSS galaxies  (about 13\%, or three times lower; see Table~\ref{tab:BPT-content}). This means that a major part of X-ray emission in galaxies comes from central SMBHs. However, SMBHs are not the only source of X-ray emission from galaxies; it can arise from star forming regions as well. When we compare the sample of $204~895$ SDSS galaxies with 1347 X-ray-selected galaxies, we see that the two samples differ in their  fraction of star-forming galaxies. More than half (57\%) of the SDSS-selected galaxies are star-forming
galaxies as defined based on their nuclear activity, while for the X-ray-selected galaxies, this fraction is less than one-quarter (23\%). We conclude that most galaxies in the HII regime have no detectable X-ray emission (weak X-ray emission from star formation may be present). We explore the
star-formation activity in a galaxy 
using the optical spectra, and the $H_{\alpha}$ line in particular; from $[OIII]$, we can obtain information about AGN activity \citet{Panessa2006}.

\begin{figure}
        
        \includegraphics[width=0.9\columnwidth]{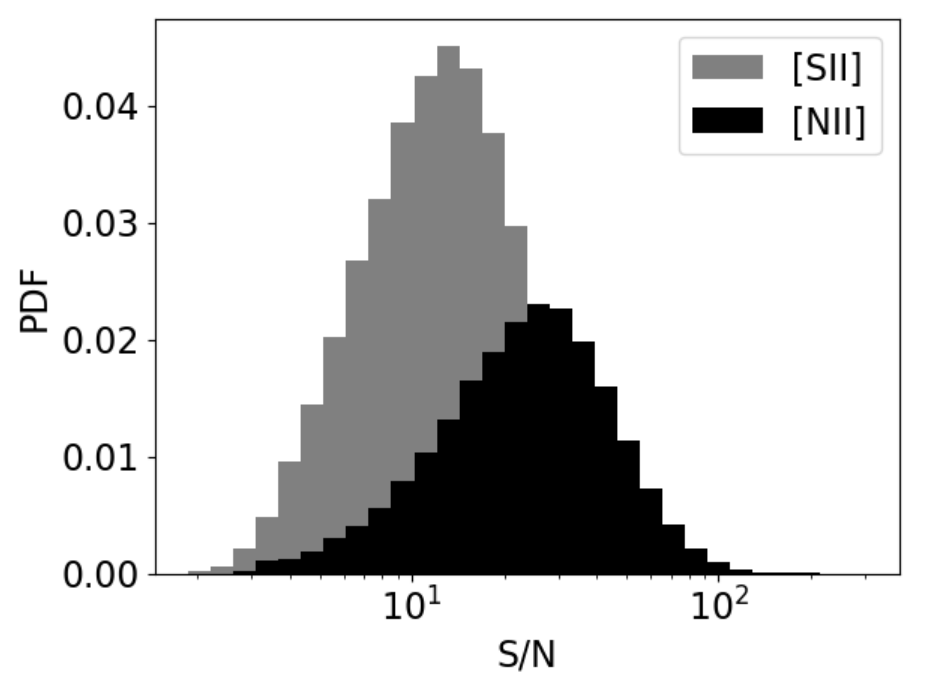}
    \caption{Probability density function 
    %\LEt{*** acronyms must be introduced in the main text only.***}
    of S/N for [NII] and [SII] doublets for 204 895 SDSS galaxy spectra. We choose the weakest line for S/N selection criteria; that is, [SII].
    }
    \label{fig:figure_err_sn}
\end{figure}

\begin{figure*}
        
        \includegraphics[width=2\columnwidth]{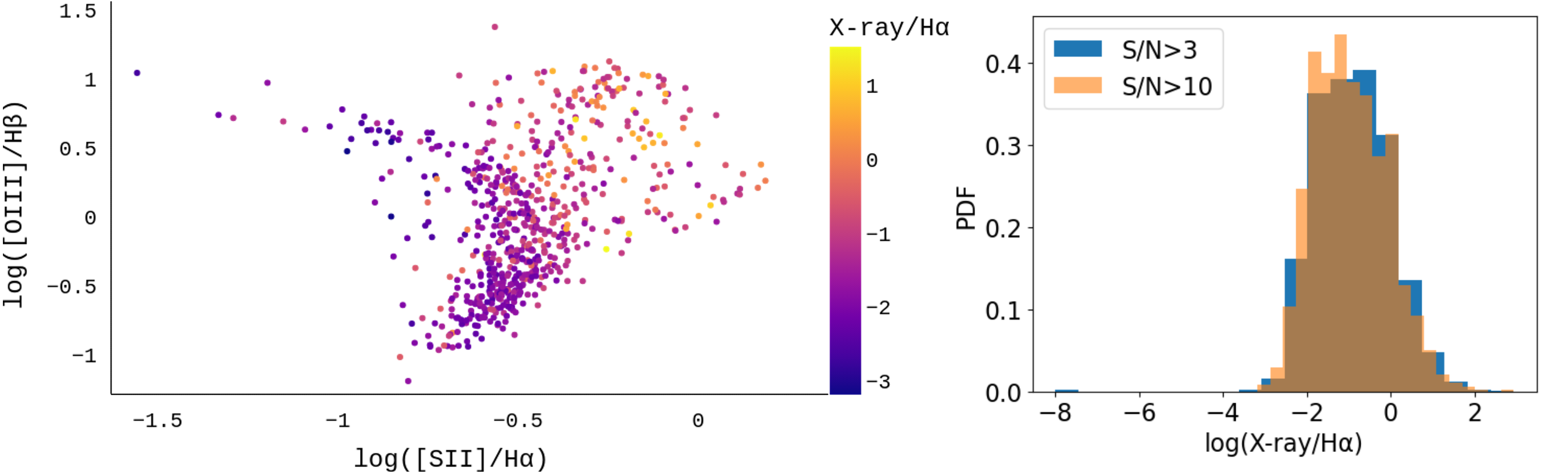}
    \caption{Distribution of SDSS X-ray-selected galaxies \textit{Left:} 
    %\LEt{***Please provide a general title before introducing the different panels. Please make sure all figures have a general title.***} 
    Galaxies with S/N>30 on the [SII] version of the BPT diagram \citep{Baldwinet1981}. Compared to Fig.~\ref{fig:BPT-DR17}, the contribution of composite-type galaxies increases. The colour bar represents the X-ray/$H\alpha$ flux ratio. \textit{Right:} PDF of $\log (X-ray/H\alpha)$ for S/N 3 and 10, respectively.
    }
    \label{fig:figure_err}
\end{figure*}

In all these analyses involving the BPT diagram, we need to define a minimum quality cut for the flux measurements of all emission lines to ensure sufficient  precision on the locations of all entries within the BPT diagram. Therefore, we chose the weakest line with the lowest signal-to-noise ratio (S/N) for the sample, namely [SII] (see Fig.~\ref{fig:figure_err_sn}). Following convention \citep[e.g.][]{Bolton2012A}, we adopt a maximal fractional flux uncertainty of 33\% or 0.13~dex (see Eq.~\ref{eq:con1} if $\frac{\sigma_{flux}}{f}<0.33,$) for the $H\alpha$, [NII], [SII], and [OIII] lines. We find that at high S/N (S/N>30, Fig.~\ref{fig:figure_err}), the fraction of composite galaxies increases significantly compared to SF and Seyfert+LINER. Therefore, increasing the S/N > 3 may influence the selection criteria, as we will get a higher fraction of composite galaxies, which are prevalent in the local Universe. The right panel of Figure~\ref{fig:figure_err}  shows that the probability density function for $\log (X-ray/H\alpha)$ has a more symmetrical form with S/N >3 compared to S/N>10. We conclude that S/N>3 is an optimal criterion that provides results of  sufficient quality and allows us to create the most homogeneous sample of objects with which to study their properties.

\begin{equation}
     \sigma_{\log f}=\frac{d \log f}{df}\cdot \sigma_{flux}=\frac{1}{\ln 10}\cdot \frac{\sigma_{flux}}{f}=\frac{0.3}{\ln 10}=0.13
        \label{eq:con1}
.\end{equation}

 Summarising our findings and conclusions, we find a clear dependence of the BPT position on the X-ray-to-$H\alpha$ flux ratio. Sources dominated by X-ray emission lie in the AGN and LINER regime of the BPT diagram. Almost all sources with low X-ray-to-H$\alpha$ luminosity ratio, $log_{10}(L_{X}/L_{H\alpha} )< 1.0$, lie in the HII regime. Our findings suggest that the X-ray-to-$H\alpha$ ratio can be used to differentiate galaxies whose X-ray flux is dominated by an AGN from those with central X-ray binaries and other stellar X-ray sources. 
We find that from XMM-Newton observations of Seyfert galaxies, the fraction of the total X-ray emission coming from the stellar component   is less than $1/30$. The galaxies with more  than 50\% of their  total X-ray emission coming from the stellar component ($log_{10}(L^{Star}_{XR}/L^{Total}_{Xray})>-0.3$ for 45 galaxies of our sample) are characterised by a dominance of X-ray emission from the stellar component. 
The galaxies for which the highest fraction  of X-ray emission is from the stellar component are  located at the bottom of the BPT diagram and are the galaxies that follow the green line that separates LINERs from Seyferts. These galaxies could be candidates for pure X-ray AGN-free galaxies.
%\LEt{***Please check that I have retained your intended meaning.***}

%-------------------------------------- Two column figure (place early!)

\begin{acknowledgements}

Pulatova Nadiia is grateful to the Deutsche Forschungsgemeinschaft for supporting the project and managing director of the Max Planck Institute for Astronomy, Thomas Henning, for his kind invitation to work at the Institute, which made this work possible. We are grateful to Igor Chilingarian, Andrea Merloni and Andrey Zhiglo for their useful comments and advice on this work. L. V. Zadorozhna is thankful to Dr. Chunshan Lin for his hospitality at the Institute of Theoretical Physics, Jagiellonian University. 
L.V. Zadorozhna gratefully acknowledges financial support from the National Science Center Poland, grants No. UMO-2021/40/C/ST9/00015, UMO-2018/30/Q/ST9/00795. 
This research was carried out with the support of the Center for the Collective Use of Scientific Equipment ”Laboratory of High Energy Physics and Astrophysics” of the Taras Shevchenko National University of Kyiv.
The authors are grateful to Alexander J. Dimoff (Max-Planck-Institut f\"ur Astronomie) for the helpful suggestion, which has improved the English text of our paper. We thank the anonymous referee for the helpful comments. 

This research has made use of data obtained from the 4XMM XMM-Newton serendipitous source catalogue compiled by the 10 institutes of the XMM-Newton Survey Science Centre selected by ESA.

Funding for the Sloan Digital Sky Survey IV has been provided by the Alfred P. Sloan Foundation, the U.S. 
Department of Energy Office of Science, and the Participating Institutions. 

SDSS-IV acknowledges support and resources from the Center for High Performance Computing  at the 
University of Utah. The SDSS website is www.sdss.org.

SDSS-IV is managed by the Astrophysical Research Consortium for the Participating Institutions 
of the SDSS Collaboration including the Brazilian Participation Group, the Carnegie Institution for Science, 
Carnegie Mellon University, Center for Astrophysics | Harvard \& Smithsonian, the Chilean Participation 
Group, the French Participation Group, Instituto de Astrof\'isica de Canarias, The Johns Hopkins 
University, Kavli Institute for the Physics and Mathematics of the Universe (IPMU) / University of 
Tokyo, the Korean Participation Group, Lawrence Berkeley National Laboratory, Leibniz Institut f\"ur Astrophysik 
Potsdam (AIP),  Max-Planck-Institut f\"ur Astronomie (MPIA Heidelberg), Max-Planck-Institut f\"ur 
Astrophysik (MPA Garching), Max-Planck-Institut f\"ur Extraterrestrische Physik (MPE), National Astronomical Observatories of 
China, New Mexico State University, New York University, University of Notre Dame, Observat\'ario 
Nacional / MCTI, The Ohio State University, Pennsylvania State University, Shanghai Astronomical Observatory, United Kingdom Participation Group, Universidad Nacional Aut\'onoma de M\'exico, University of Arizona, University of Colorado Boulder, University of Oxford, University of Portsmouth, University of Utah, University of Virginia, University of Washington, University of Wisconsin, Vanderbilt University, and Yale University.

%%%%%%%%%%%%%%%%%%%%%%%%%%%%%%%%%%%%%%%%%%%%%%%%%%
\\
\textbf{Data availability}
\\
 
All the data presented and used in this paper are publicly available. The SDSS data (DR17 release) are accessible through the online SQL search service on the web  page at 
http://skyserver.sdss.org/dr17/SearchTools/sql. 
The XMM-Newton data are available in the XMM-Newton Survey
Science Centre. 
%(http://xmmssc.irap.omp.eu/Catalogue/3XMM-DR8/3XMM_DR8.html). 

\end{acknowledgements}

\begin{thebibliography}{31}
\expandafter\ifx\csname natexlab\endcsname\relax\def\natexlab#1{#1}\fi

\bibitem[{{Ananna} {et~al.}(2017){Ananna}, {Salvato}, {LaMassa}, {Urry},
  {Cappelluti}, {Cardamone}, {Civano}, {Farrah}, {Gilfanov}, {Glikman},
  {Hamilton}, {Kirkpatrick}, {Lanzuisi}, {Marchesi}, {Merloni}, {Nandra},
  {Natarajan}, {Richards}, \& {Timlin}}]{Ananna2017}
{Ananna}, T.~T., {Salvato}, M., {LaMassa}, S., {et~al.} 2017, \apj, 850, 66

\bibitem[{{Baldwin} {et~al.}(1981){Baldwin}, {Phillips}, \&
  {Terlevich}}]{Baldwinet1981}
{Baldwin}, J.~A., {Phillips}, M.~M., \& {Terlevich}, R. 1981, \pasp, 93, 5

\bibitem[{{Birchall} {et~al.}(2022){Birchall}, {Watson}, {Aird}, \&
  {Starling}}]{Birchallet2022}
{Birchall}, K.~L., {Watson}, M.~G., {Aird}, J., \& {Starling}, R.~L.~C. 2022,
  \mnras, 510, 4556

\bibitem[{{Blanton} {et~al.}(2017){Blanton}, {Bershady}, {Abolfathi},
  {Albareti}, {Allende Prieto}, {Almeida}, {Alonso-Garc{\'\i}a}, {Anders},
  {Anderson}, {Andrews}, {Aquino-Ort{\'\i}z}, {Arag{\'o}n-Salamanca},
  {Argudo-Fern{\'a}ndez}, {Armengaud}, {Aubourg}, {Avila-Reese}, {Badenes},
  {Bailey}, {Barger}, {Barrera-Ballesteros}, {Bartosz}, {Bates}, {Baumgarten},
  {Bautista}, {Beaton}, {Beers}, {Belfiore}, {Bender}, {Berlind}, {Bernardi},
  {Beutler}, {Bird}, {Bizyaev}, {Blanc}, {Blomqvist}, {Bolton}, {Boquien},
  {Borissova}, {van den Bosch}, {Bovy}, {Brandt}, {Brinkmann}, {Brownstein},
  {Bundy}, {Burgasser}, {Burtin}, {Busca}, {Cappellari}, {Delgado Carigi},
  {Carlberg}, {Carnero Rosell}, {Carrera}, {Chanover}, {Cherinka}, {Cheung},
  {G{\'o}mez Maqueo Chew}, {Chiappini}, {Choi}, {Chojnowski}, {Chuang},
  {Chung}, {Cirolini}, {Clerc}, {Cohen}, {Comparat}, {da Costa}, {Cousinou},
  {Covey}, {Crane}, {Croft}, {Cruz-Gonzalez}, {Garrido Cuadra}, {Cunha},
  {Damke}, {Darling}, {Davies}, {Dawson}, {de la Macorra}, {Dell'Agli}, {De
  Lee}, {Delubac}, {Di Mille}, {Diamond-Stanic}, {Cano-D{\'\i}az}, {Donor},
  {Downes}, {Drory}, {du Mas des Bourboux}, {Duckworth}, {Dwelly}, {Dyer},
  {Ebelke}, {Eigenbrot}, {Eisenstein}, {Emsellem}, {Eracleous}, {Escoffier},
  {Evans}, {Fan}, {Fern{\'a}ndez-Alvar}, {Fernandez-Trincado}, {Feuillet},
  {Finoguenov}, {Fleming}, {Font-Ribera}, {Fredrickson}, {Freischlad},
  {Frinchaboy}, {Fuentes}, {Galbany}, {Garcia-Dias},
  {Garc{\'\i}a-Hern{\'a}ndez}, {Gaulme}, {Geisler}, {Gelfand},
  {Gil-Mar{\'\i}n}, {Gillespie}, {Goddard}, {Gonzalez-Perez}, {Grabowski},
  {Green}, {Grier}, {Gunn}, {Guo}, {Guy}, {Hagen}, {Hahn}, {Hall}, {Harding},
  {Hasselquist}, {Hawley}, {Hearty}, {Gonzalez Hern{\'a}ndez}, {Ho}, {Hogg},
  {Holley-Bockelmann}, {Holtzman}, {Holzer}, {Huehnerhoff}, {Hutchinson},
  {Hwang}, {Ibarra-Medel}, {da Silva Ilha}, {Ivans}, {Ivory}, {Jackson},
  {Jensen}, {Johnson}, {Jones}, {J{\"o}nsson}, {Jullo}, {Kamble}, {Kinemuchi},
  {Kirkby}, {Kitaura}, {Klaene}, {Knapp}, {Kneib}, {Kollmeier}, {Lacerna},
  {Lane}, {Lang}, {Law}, {Lazarz}, {Lee}, {Le Goff}, {Liang}, {Li}, {Li},
  {Lian}, {Lima}, {Lin}, {Lin}, {Bertran de Lis}, {Liu}, {de Icaza Lizaola},
  {Long}, {Lucatello}, {Lundgren}, {MacDonald}, {Deconto Machado}, {MacLeod},
  {Mahadevan}, {Geimba Maia}, {Maiolino}, {Majewski}, {Malanushenko},
  {Malanushenko}, {Manchado}, {Mao}, {Maraston}, {Marques-Chaves}, {Masseron},
  {Masters}, {McBride}, {McDermid}, {McGrath}, {McGreer}, {Medina Pe{\~n}a},
  {Melendez}, {Merloni}, {Merrifield}, {Meszaros}, {Meza}, {Minchev},
  {Minniti}, {Miyaji}, {More}, {Mulchaey}, {M{\"u}ller-S{\'a}nchez}, {Muna},
  {Munoz}, {Myers}, {Nair}, {Nandra}, {Correa do Nascimento}, {Negrete},
  {Ness}, {Newman}, {Nichol}, {Nidever}, {Nitschelm}, {Ntelis}, {O'Connell},
  {Oelkers}, {Oravetz}, {Oravetz}, {Pace}, {Padilla}, {Palanque-Delabrouille},
  {Alonso Palicio}, {Pan}, {Parejko}, {Parikh}, {P{\^a}ris}, {Park}, {Patten},
  {Peirani}, {Pellejero-Ibanez}, {Penny}, {Percival}, {Perez-Fournon},
  {Petitjean}, {Pieri}, {Pinsonneault}, {Pisani}, {Poleski}, {Prada},
  {Prakash}, {Queiroz}, {Raddick}, {Raichoor}, {Barboza Rembold}, {Richstein},
  {Riffel}, {Riffel}, {Rix}, {Robin}, {Rockosi}, {Rodr{\'\i}guez-Torres},
  {Roman-Lopes}, {Rom{\'a}n-Z{\'u}{\~n}iga}, {Rosado}, {Ross}, {Rossi}, {Ruan},
  {Ruggeri}, {Rykoff}, {Salazar-Albornoz}, {Salvato}, {S{\'a}nchez}, {Aguado},
  {S{\'a}nchez-Gallego}, {Santana}, {Santiago}, {Sayres}, {Schiavon}, {da Silva
  Schimoia}, {Schlafly}, {Schlegel}, {Schneider}, {Schultheis}, {Schuster},
  {Schwope}, {Seo}, {Shao}, {Shen}, {Shetrone}, {Shull}, {Simon}, {Skinner},
  {Skrutskie}, {Slosar}, {Smith}, {Sobeck}, {Sobreira}, {Somers}, {Souto},
  {Stark}, {Stassun}, {Stauffer}, {Steinmetz}, {Storchi-Bergmann},
  {Streblyanska}, {Stringfellow}, {Su{\'a}rez}, {Sun}, {Suzuki}, {Szigeti},
  {Taghizadeh-Popp}, {Tang}, {Tao}, {Tayar}, {Tembe}, {Teske}, {Thakar},
  {Thomas}, {Thompson}, {Tinker}, {Tissera}, {Tojeiro}, {Hernandez Toledo}, {de
  la Torre}, {Tremonti}, {Troup}, {Valenzuela}, {Martinez Valpuesta},
  {Vargas-Gonz{\'a}lez}, {Vargas-Maga{\~n}a}, {Vazquez}, {Villanova}, {Vivek},
  {Vogt}, {Wake}, {Walterbos}, {Wang}, {Weaver}, {Weijmans}, {Weinberg},
  {Westfall}, {Whelan}, {Wild}, {Wilson}, {Wood-Vasey}, {Wylezalek}, {Xiao},
  {Yan}, {Yang}, {Ybarra}, {Y{\`e}che}, {Zakamska}, {Zamora}, {Zarrouk},
  {Zasowski}, {Zhang}, {Zhao}, {Zheng}, {Zheng}, {Zhou}, {Zhou}, {Zhu},
  {Zoccali}, \& {Zou}}]{2017AJ....154...28B}
{Blanton}, M.~R., {Bershady}, M.~A., {Abolfathi}, B., {et~al.} 2017, \aj, 154,
  28

\bibitem[{{Bolton} {et~al.}(2012){Bolton}, {Schlegel}, {Aubourg}, {Bailey},
  {Bhardwaj}, {Brownstein}, {Burles}, {Chen}, {Dawson}, {Eisenstein}, {Gunn},
  {Knapp}, {Loomis}, {Lupton}, {Maraston}, {Muna}, {Myers}, {Olmstead},
  {Padmanabhan}, {P{\^a}ris}, {Percival}, {Petitjean}, {Rockosi}, {Ross},
  {Schneider}, {Shu}, {Strauss}, {Thomas}, {Tremonti}, {Wake}, {Weaver}, \&
  {Wood-Vasey}}]{Bolton2012A}
{Bolton}, A.~S., {Schlegel}, D.~J., {Aubourg}, {\'E}., {et~al.} 2012, \aj, 144,
  144

\bibitem[{{Brandt} \& {Hasinger}(2005)}]{Brandt2005}
{Brandt}, W.~N. \& {Hasinger}, G. 2005, \araa, 43, 827

\bibitem[{{Brescia} {et~al.}(2015){Brescia}, {Cavuoti}, \&
  {Longo}}]{Bresciaetal2015}
{Brescia}, M., {Cavuoti}, S., \& {Longo}, G. 2015, \mnras, 450, 3893

\bibitem[{{Chang} {et~al.}(2015){Chang}, {van der Wel}, {da Cunha}, \&
  {Rix}}]{Chang2015}
{Chang}, Y.-Y., {van der Wel}, A., {da Cunha}, E., \& {Rix}, H.-W. 2015, \apjs,
  219, 8

\bibitem[{{Costa} {et~al.}(2014){Costa}, {Sijacki}, {Trenti}, \&
  {Haehnelt}}]{Costa2014}
{Costa}, T., {Sijacki}, D., {Trenti}, M., \& {Haehnelt}, M.~G. 2014, \mnras,
  439, 2146

\bibitem[{{Cresci} \& {Maiolino}(2018)}]{Cresci2018}
{Cresci}, G. \& {Maiolino}, R. 2018, Nature Astronomy, 2, 179

\bibitem[{{Croton} {et~al.}(2006){Croton}, {Springel}, {White}, {De Lucia},
  {Frenk}, {Gao}, {Jenkins}, {Kauffmann}, {Navarro}, \& {Yoshida}}]{Croton2006}
{Croton}, D.~J., {Springel}, V., {White}, S. D.~M., {et~al.} 2006, \mnras, 365,
  11

\bibitem[{{Elvis} {et~al.}(1984){Elvis}, {Soltan}, \& {Keel}}]{Elvis1984}
{Elvis}, M., {Soltan}, A., \& {Keel}, W.~C. 1984, \apj, 283, 479

\bibitem[{{Fitriana} \& {Murayama}(2022)}]{Fitriana2022}
{Fitriana}, I.~K. \& {Murayama}, T. 2022, \pasj [\eprint[arXiv]{2203.15967}]

\bibitem[{{Hopkins} {et~al.}(2016){Hopkins}, {Torrey}, {Faucher-Gigu{\`e}re},
  {Quataert}, \& {Murray}}]{Hopkins2016}
{Hopkins}, P.~F., {Torrey}, P., {Faucher-Gigu{\`e}re}, C.-A., {Quataert}, E.,
  \& {Murray}, N. 2016, \mnras, 458, 816

\bibitem[{{Ishibashi} \& {Fabian}(2012)}]{Ishibashi2012}
{Ishibashi}, W. \& {Fabian}, A.~C. 2012, \mnras, 427, 2998

\bibitem[{{Kauffmann} {et~al.}(2003){Kauffmann}, {Heckman}, {Tremonti},
  {Brinchmann}, {Charlot}, {White}, {Ridgway}, {Brinkmann}, {Fukugita}, {Hall},
  {Ivezi{\'c}}, {Richards}, \& {Schneider}}]{Kauffmann2003}
{Kauffmann}, G., {Heckman}, T.~M., {Tremonti}, C., {et~al.} 2003, \mnras, 346,
  1055

\bibitem[{{Kewley} {et~al.}(2006){Kewley}, {Groves}, {Kauffmann}, \&
  {Heckman}}]{Kewleyet2006}
{Kewley}, L.~J., {Groves}, B., {Kauffmann}, G., \& {Heckman}, T. 2006, \mnras,
  372, 961

\bibitem[{{LaMassa} {et~al.}(2016){LaMassa}, {Urry}, {Cappelluti},
  {B{\"o}hringer}, {Comastri}, {Glikman}, {Richards}, {Ananna}, {Brusa},
  {Cardamone}, {Chon}, {Civano}, {Farrah}, {Gilfanov}, {Green}, {Komossa},
  {Lira}, {Makler}, {Marchesi}, {Pecoraro}, {Ranalli}, {Salvato}, {Schawinski},
  {Stern}, {Treister}, \& {Viero}}]{LaMassa2016}
{LaMassa}, S.~M., {Urry}, C.~M., {Cappelluti}, N., {et~al.} 2016, \apj, 817,
  172

\bibitem[{{LaMassa} {et~al.}(2013){LaMassa}, {Urry}, {Cappelluti}, {Civano},
  {Ranalli}, {Glikman}, {Treister}, {Richards}, {Ballantyne}, {Stern},
  {Comastri}, {Cardamone}, {Schawinski}, {B{\"o}hringer}, {Chon}, {Murray},
  {Green}, \& {Nandra}}]{LaMassa2013}
{LaMassa}, S.~M., {Urry}, C.~M., {Cappelluti}, N., {et~al.} 2013, \mnras, 436,
  3581

\bibitem[{{Lehmer} {et~al.}(2010){Lehmer}, {Alexander}, {Bauer}, {Brandt},
  {Goulding}, {Jenkins}, {Ptak}, \& {Roberts}}]{Lehmer2010}
{Lehmer}, B.~D., {Alexander}, D.~M., {Bauer}, F.~E., {et~al.} 2010, \apj, 724,
  559

\bibitem[{{Mulchaey} {et~al.}(1994){Mulchaey}, {Koratkar}, {Ward}, {Wilson},
  {Whittle}, {Antonucci}, {Kinney}, \& {Hurt}}]{Mulchaey1994}
{Mulchaey}, J.~S., {Koratkar}, A., {Ward}, M.~J., {et~al.} 1994, \apj, 436, 586

\bibitem[{{Panessa} {et~al.}(2006){Panessa}, {Bassani}, {Cappi}, {Dadina},
  {Barcons}, {Carrera}, {Ho}, \& {Iwasawa}}]{Panessa2006}
{Panessa}, F., {Bassani}, L., {Cappi}, M., {et~al.} 2006, \aap, 455, 173

\bibitem[{{Schartmann} {et~al.}(2009){Schartmann}, {Meisenheimer}, {Klahr},
  {Camenzind}, {Wolf}, \& {Henning}}]{Schartmann2009}
{Schartmann}, M., {Meisenheimer}, K., {Klahr}, H., {et~al.} 2009, \mnras, 393,
  759

\bibitem[{{Shangguan} {et~al.}(2020){Shangguan}, {Ho}, {Bauer}, {Wang}, \&
  {Treister}}]{Shangguan2020}
{Shangguan}, J., {Ho}, L.~C., {Bauer}, F.~E., {Wang}, R., \& {Treister}, E.
  2020, \apj, 899, 112

\bibitem[{Stemo {et~al.}(2020)Stemo, Comerford, Barrows, Stern, Assef, \&
  Griffith}]{Stemo2020}
Stemo, A., Comerford, J.~M., Barrows, R.~S., {et~al.} 2020, The Astrophysical
  Journal, 888, 78

\bibitem[{{Str{\"u}der} {et~al.}(2001){Str{\"u}der}, {Briel}, {Dennerl},
  {Hartmann}, {Kendziorra}, {Meidinger}, {Pfeffermann}, {Reppin}, {Aschenbach},
  {Bornemann}, {Br{\"a}uninger}, {Burkert}, {Elender}, {Freyberg}, {Haberl},
  {Hartner}, {Heuschmann}, {Hippmann}, {Kastelic}, {Kemmer}, {Kettenring},
  {Kink}, {Krause}, {M{\"u}ller}, {Oppitz}, {Pietsch}, {Popp}, {Predehl},
  {Read}, {Stephan}, {St{\"o}tter}, {Tr{\"u}mper}, {Holl}, {Kemmer}, {Soltau},
  {St{\"o}tter}, {Weber}, {Weichert}, {von Zanthier}, {Carathanassis}, {Lutz},
  {Richter}, {Solc}, {B{\"o}ttcher}, {Kuster}, {Staubert}, {Abbey}, {Holland},
  {Turner}, {Balasini}, {Bignami}, {La Palombara}, {Villa}, {Buttler},
  {Gianini}, {Lain{\'e}}, {Lumb}, \& {Dhez}}]{Struder2001}
{Str{\"u}der}, L., {Briel}, U., {Dennerl}, K., {et~al.} 2001, \aap, 365, L18

\bibitem[{{Torbaniuk} {et~al.}(2021){Torbaniuk}, {Paolillo}, {Carrera},
  {Cavuoti}, {Vignali}, {Longo}, \& {Aird}}]{Torbaniuk2021}
{Torbaniuk}, O., {Paolillo}, M., {Carrera}, F., {et~al.} 2021, \mnras, 506,
  2619

\bibitem[{{Tugay} \& {Vasylenko}(2011)}]{Tugay2011}
{Tugay}, A.~V. \& {Vasylenko}, A.~A. 2011, Odessa Astronomical Publications,
  24, 72

\bibitem[{{Ward} {et~al.}(1988){Ward}, {Done}, {Fabian}, {Tennant}, \&
  {Shafer}}]{Ward1988}
{Ward}, M.~J., {Done}, C., {Fabian}, A.~C., {Tennant}, A.~F., \& {Shafer},
  R.~A. 1988, \apj, 324, 767

\bibitem[{{Webb} {et~al.}(2020){Webb}, {Coriat}, {Traulsen}, {Ballet}, {Motch},
  {Carrera}, {Koliopanos}, {Authier}, {de la Calle}, {Ceballos}, {Colomo},
  {Chuard}, {Freyberg}, {Garcia}, {Kolehmainen}, {Lamer}, {Lin}, {Maggi},
  {Michel}, {Page}, {Page}, {Perea-Calderon}, {Pineau}, {Rodriguez}, {Rosen},
  {Santos Lleo}, {Saxton}, {Schwope}, {Tom{\'a}s}, {Watson}, \&
  {Zakardjian}}]{webb20}
{Webb}, N.~A., {Coriat}, M., {Traulsen}, I., {et~al.} 2020, \aap, 641, A136

\bibitem[{{Zadorozhna} {et~al.}(2021){Zadorozhna}, {Tugay}, {Shevchenko}, \&
  {Pulatova}}]{zador21}
{Zadorozhna}, L.~V., {Tugay}, A.~V., {Shevchenko}, S.~Y., \& {Pulatova}, N.~G.
  2021, Kinematics and Physics of Celestial Bodies, 37, 149

%\end{thebibliography}
%\bibliography{main}
\end{thebibliography}

\bibliographystyle{aa}
%\begin{thebibliography}

%%%%%%%%%%%%%%%%% APPENDICES %%%%%%%%%%%%%%%%%%%%%

\begin{appendix}
%\appendix
\section{Rejected objects} 
\label{append}

\renewcommand{\thefigure}{A.\arabic{figure}} % Change figure numbering
\setcounter{figure}{0} % Reset figure counter
\renewcommand{\thetable}{A.\arabic{table}} % Change table numbering
\setcounter{table}{0} % Reset table counter

%\FloatBarrier 

\begin{figure}[ht]
        \includegraphics[width=\columnwidth]{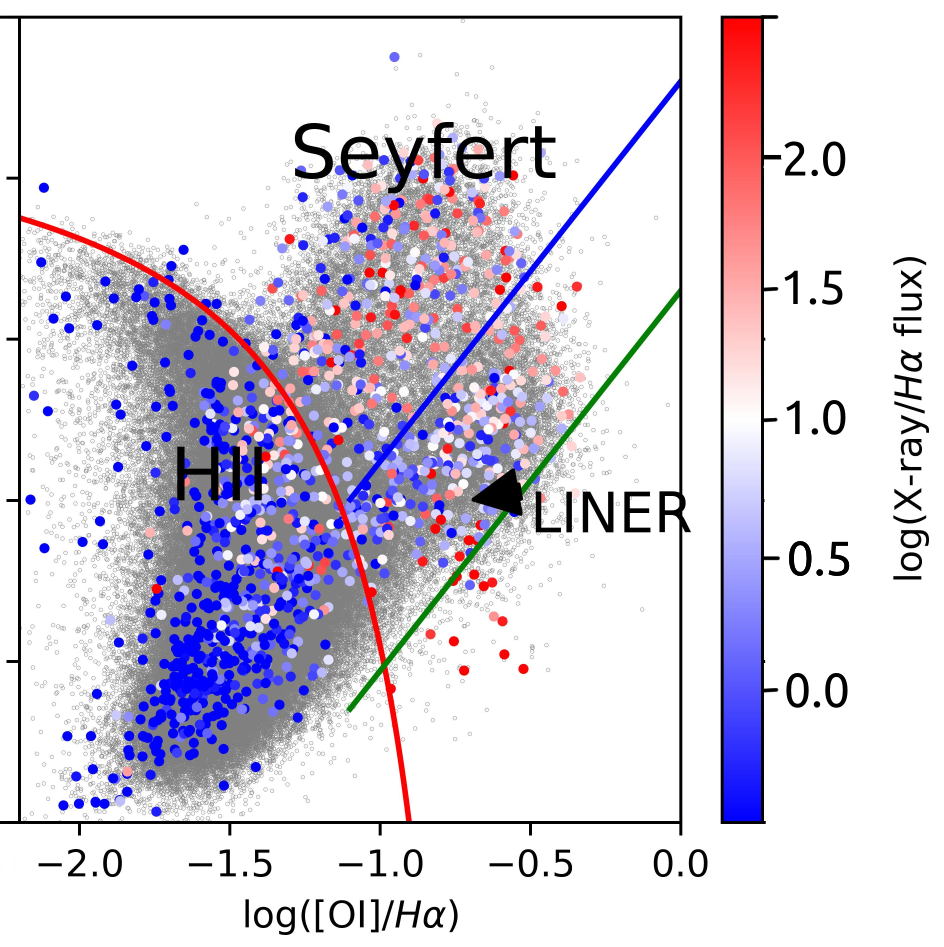}
        \caption{The position on $[OIII]/H_\beta$ vs $[OI]/H_\alpha$ BPT diagram of selected SDSS galaxies with X-ray nuclear emission. Empirical lines (dashed blue and solid red) that separate LINERs from Seyfert galaxies are from \citet{Kewleyet2006}. The green empirical line ($[OIII]/H_\beta < 0.64 + 1.18 \cdot [OI]/H_\alpha$) separates "true" LINERs from "doubtful" galaxies.  }
        \label{fig:False-gal-BPT}
\end{figure}

\begin{table*}[ht]
        \centering
        \caption{Information about 19 individual SDSS galaxies with nuclear X-ray emission, located under LINER position on BPT diagram}
        \label{tab:doubtful-galaxies}
        \begin{tabular}{lcccr} % four columns, alignment for each
                \hline
                Number & RA & DER & SDSS classification & remarks \\
                \hline
        1       &       16.70573        &       1.05625 &       STARBURST BROADLINE       &               \\
        2       &       28.28456        &       1.03075 &       STARFORMING     &               \\
        3       &       117.85459       &       17.51422        &       STARFORMING     &       absent  $H_{\beta}$    \\
        4       &       121.63670       &       15.26223        &       ------- &       absent  $H_{\beta}$    \\
        5       &       155.91514       &       4.18629 &       STARBURST BROADLINE       &               \\
        6       &       167.80496       &       28.69641        &       AGN BROADLINE       &               \\
        7       &       168.96626       &       1.49863 &       STARFORMING     &       absent [OI], S/N = 4   \\
        8       &       179.03689       &       56.77642        &       ------- &       absent  $H_{\beta}$, [OIII]    \\
        9       &       183.24640       &       27.45127        &       STARFORMING     &       absent [OIII], [OI]    \\
        10      &       195.92004       &       53.72324        &       STARFORMING     &               \\
        11      &       202.86815       &       25.06823        &       ------- &       absent  emission lines \\
        12      &       215.99761       &       -0.89128        &       AGN     &       absent  $H_{\beta}$, [OIII], [OI]      \\
        13      &       224.31295       &       22.34289        &       STARFORMING BROADLINE       &       emission lines has double peaks         \\
        14      &       233.22408       &       30.34984        &       STARBURST       &       $\sigma_{line} \sim~200~km/s$  \\
        15      &       243.92172       &       47.18661        &       ------- &       absent  $H_{\beta}$, [OIII], [OI]      \\
        16      &       260.04183       &       26.62557        &       STARFORMING BROADLINE       &       absent [OIII]   \\
        17      &       322.41647       &       0.08921 &       STARFORMING     &       absent [OIII]  \\
        18      &       342.90562       &       1.19148 &       ----------      &       absent  $H_{\beta}$    \\
        19      &       358.52933       &       -10.42132       &       AGN BROADLINE       &       absent  $H_{\beta}$     \\
                \hline
        \end{tabular}
\end{table*}

We checked all 19 X-ray-selected SDSS galaxies below the LINER's position on the BPT diagram. The empirical  separation line between LINERs and `doubtful galaxies' marked with the green line, $[OIII]/H_\beta < 0.64 + 1.18 \cdot [OI]/H_\alpha$ in Fig.~\ref{fig:False-gal-BPT}. 
We find that 12 of all 19 galaxies have no emission lines that we use to plot BPT diagrams (see Table~\ref{tab:doubtful-galaxies}). The $H_\beta$ line was mostly falsely detected, and its flux was calculated automatically. Visually, this region appears to be a random fluctuation of noise. We removed these galaxies from the sample.

The other four galaxies are classified automatically by SDSS \citep{Bresciaetal2015} as BROADLINE galaxies (with $\sigma_{line} >200~km/s$). We removed these galaxies from the sample as the physical conditions that give rise to the broad lines are different from those that give rise to the narrow lines, and so comparing these types of objects is misleading.

Finally, the remaining three galaxies were classified as STARBURST or STARFORMING. The difference between the STARBURST and STARFORMING classifications of galaxies is worthy of note. A\ STARBURST galaxy is a STARFORMING galaxy with an equivalent width of $H\alpha$ of greater than $50$ \AA. From these three STARBURST or STARFORMING galaxies (numbers 2, 10, 14 in Table~\ref{tab:doubtful-galaxies}) we find that the last one, 14, has $\sigma_{line} >200~km/s$ for all-optical emission lines, except $H\alpha$, $H\beta$ with $\sigma_{line} \sim 200~km/s$ ). According to this, we decided to remove galaxy 14 from Table~\ref{tab:doubtful-galaxies} as a BROADLINE galaxy. The last two  STARFORMING galaxies (SDSS J015308.29+010150.6 and SDSS J130340.81+534323.6) are located at the bottom end of the LINER region and were not investigated further.

\end{appendix}

\end{document}